\documentclass[english,11pt,oneside]{article}
\pdfoutput=1
\usepackage{amsmath}
\usepackage{amsfonts}
\usepackage{amssymb}
\usepackage{graphicx, rotating}
\usepackage{epstopdf}
\usepackage{epsfig}
\usepackage{latexsym}
\usepackage{multirow}
\usepackage{color}
\usepackage{slashed}
\usepackage{pifont}
\usepackage[top=2.8 cm, bottom=2.8 cm, left=3 cm, right=3 cm]{geometry}
\usepackage[compat=1.0.0]{tikz-feynman}
\usepackage{mathtools}

\newcommand{\ba}{\begin{eqnarray}}
\newcommand{\ea}{\end{eqnarray}}
\newcommand{\no}{\nonumber}
\newcommand{\beal}{\begin{align}}
\newcommand{\eeal}{\end{align}}

\usepackage{hyperref}
\hypersetup{
 colorlinks = true,
 linkcolor  = red,
 citecolor={blue}
}

\usepackage{bbold}

\begin{document}

\title{
A strange contribution to the neutron EDM
}
\author{Luca Vecchi~\footnote{luca.vecchi@pd.infn.it}\\
{\small\emph{Istituto Nazionale di Fisica Nucleare (INFN), Sezione di Padova, Italy}}}
\date{}
\maketitle

\begin{abstract}

We analyze the contribution of hypothetical quark electric dipoles to the electric dipole moment (EDM) of the neutron. Particular emphasis is devoted to the strange quark contribution. Considerations based on perturbative QCD, the large N expansion, a critic reassessment of the non-relativistic quark model as well as a next to leading order calculation in heavy baryon effective field theory, all consistently indicate that, barring accidental cancellations, the matrix element of the strange quark dipole should be of order a tenth of those of the valence quarks. This implies that the strange EDM provides the dominant contribution to the neutron EDM in many scenarios beyond the Standard Model.

\end{abstract}

\newpage

{	\hypersetup{linkcolor=black}	\tableofcontents}

\section{Motivation}

The non-observation of the electric dipole moment (EDM) of the neutron, $d_n$, provides one of the most stringent constraints on physics beyond the Standard Model. Currently the best bound is $|d_n|\leq1.8\times10^{-26}~e\,{\text{cm}}$ at $90\%$ CL \cite{Abel:2020pzs}, but the experimental sensitivity is expected to improve by one or even two orders of magnitude in the next ten to twenty years \cite{Alarcon:2022ero}.

Given the enormous impact that this measurement has, it is crucial to correctly identify the mapping between the new physics parameters and $d_n$. Among the many contributions to $d_n$, those arising from hypothetical quark EDMs are especially relevant for a vast class of new physics scenarios. Denoting the EDMs of the up, down, and strange quarks renormalized slightly above the GeV as $d_u, d_d, d_s$, respectively, and working at leading order in the CP-violating couplings, the neutron EDM takes the form 
\ba\label{nEDM}
d_n=g_{Tn}^ud_{u}+g_{Tn}^dd_{d}+g_{Tn}^sd_{s}
\ea
for some unknown real numbers $g_{Tn}^{u,d,s}$ known as tensor charges.

The tensor charges are intrinsically non-perturbative and cannot be reliably computed analytically. Yet, intuitively one expects sizable $g_{Tn}^{u,d}$ and a moderately smaller $g_{Tn}^{s}$. The strange quark contributes non-negligibly to the nucleon masses (as confirmed by analyses in chiral perturbation theory \cite{Borasoy:1996bx} and lattice QCD \cite{FlavourLatticeAveragingGroupFLAG:2024oxs}), as well as to the nucleon spin (see \cite{Savage:1996zd} for chiral perturbation theory and again \cite{FlavourLatticeAveragingGroupFLAG:2024oxs} for lattice QCD measurements). The matrix elements of both the scalar and axial-vector strange operators are roughly a fraction $0.1$ of the analogous valence quark operators. One would thus naively anticipate a similar trend in the tensor charges. 

However, this is not what more concrete approaches find. As we review in the next section, the tensor charges have been estimated via various models of the QCD dynamics as well as via numerical simulations with lattice QCD methods. The former seem to suggest $g_{Tn}^{s}$ is negligible but, admittedly, either completely ignore the effect of the strange EDM or suffer from uncertainties that are impossible to quantify. Lattice QCD simulations instead measure a $|g_{Tn}^s|$ smaller by a factor of 100 to 1000 compared to the matrix elements of the valence quarks \cite{Bhattacharya:2016zcn,Gupta:2018lvp,Park:2025rxi}. At present the origin of such surprising small ratio is unknown. 

This state of affairs is not fully satisfactory, in our opinion. What must absolutely be avoided is that the actual impact of the strange EDM gets erroneously underestimated, since that would badly misguide us when assessing the phenomenological implications of the experimental efforts. Indeed, in many well-motivated scenarios of new physics the quark dipoles arise at 1-loop and the chirality-flip is controlled by parameters that feature hierarchies similar to those of the quark masses, so that $d_q\sim m_qe/m_*^2$ for some heavy mass scale $m_*$. In those scenarios the strange quark dipole would dominate over those of the up or down as soon as $|g_{Tn}^s/g_{Tn}^{u,d}|\gtrsim m_{u,d}/m_s\sim0.02\div0.05$. And in the extreme event in which $|g_{Tn}^s/g_{Tn}^{u,d}|\sim1$ the lower bounds on the new physics scale would be stronger by a factor of $4$ to $6$ compared to those expected from valence quark dipoles. 
Evidently, which new physics scenarios are or are not allowed by a measurement of $d_n$ critically depends on the value of $g_{Tn}^s$, and qualitatively differ if, say, $|g_{Tn}^s/g_{Tn}^{u,d}|\sim0.1$ as opposed to $|g_{Tn}^s/g_{Tn}^{u,d}|\sim0.01$.

The relevance of the sea quarks in nucleon physics has a very long history, and has been and is being investigated by many authors. In the study of the neutron EDM, the impact of the strange {\emph{chromo-electric}} dipole was emphasized by several authors in the past (see e.g. \cite{Khatsimovsky:1987bb,Hisano:2004tf,Fuyuto:2012yf}). Surprisingly, though, we have not found a likewise systematic assessment of the strange EDM contribution to $d_n$. The main aim of the present paper is filling this gap. While the last word is clearly left to the lattice QCD community, we believe that the transparency and simplicity of an analytical approach can, despite all its limits, help shed light on this interesting problem.

The paper is organized as follows. We begin with a presentation of the setup in Section \ref{sec:pheno}. The most popular estimates of the tensor charges are summarized in Section \ref{sec:tensor}. A first look at the diagrams contributing to $g_{Tn}^s$ is presented in Section \ref{sec:RG}. 
In Section \ref{sec:largeN} we complete our preliminary assessment by showing what large $N_c$ QCD has to say about the tensor charges. In Section \ref{sec:NRQM} we start a more quantitative analysis, critically reviewing the predictions of the familiar non-relativistic quark model. In Section \ref{sec:chiralPT} we present the result of a chiral expansion at next to leading order. The appendices contain details on the results presented in the bulk of the paper. We summarize our conclusions in Section \ref{sec:conclusions}.

\section{Introductory remarks}
\label{sec:pheno}

We perturb the QCD Lagrangian introducing electric dipole moments for the light quarks. At a scale $\Lambda_{\text{UV}}$ above the typical hadron mass $M_h\sim1$ GeV the Lagrangian reads
\begin{align}\label{Lag}
{\cal L}_{\Lambda_{\text{UV}}}
&=-\frac{1}{4}G_{\mu\nu}^AG^{A\,\mu\nu}+\overline{q_i}i{\slashed{D}}q_i-(\overline{q_{i}}m_{ij}P_Rq_{j}+{\text{hc}})-\theta\frac{g^2}{64\pi^2}G_{\mu\nu}^AG_{\alpha\beta}^A\epsilon^{\mu\nu\alpha\beta}\\\no
&-\left(\frac{d_{ij}}{2}\overline{q}_ii\sigma^{\mu\nu}P_Rq_j\,F_{\mu\nu}+{\text{hc}}\right),
\end{align}
where a sum over the flavor indices $i,j=u,d,s$ is understood, $P_R=\frac12(1+\gamma_5)$ is the right-handed chiral projector, and finally $\sigma^{\mu\nu}=\frac{i}{2}[\gamma^\mu,\gamma^\nu]$. For simplicity, throughout the paper we will work in a fantasy world in which the charm quark is much heavier than $M_h$. Hence, at the scales of interest, the $c,b,t$ quarks are sufficiently heavy to be integrated out and their impact on $d_n$ is suppressed by powers of the heavy quark mass. In practice this means that
\ba\label{LambdaUV}
M_h<\Lambda_{\text{UV}}<m_c.
\ea
So, to respect the assumed hierarchy, in the real world $\Lambda_{\text{UV}}$ should be rather close to the non-perturbative scale of QCD. 

We have experimental evidence that vacuum alignment in real-world QCD is determined approximately by $m$. Formally this means that nature is well approximated by the assumption that the absolute value of $\bar\theta\equiv\theta+{\text{Arg}}{\text{Det}}[m]$ is negligible and that the dipoles are small in units of the hadron scale, i.e. ${|d|e M_h}/{(4\pi)^2}\ll |m|/{M_h}$. It is convenient to work with a field basis in which the quark mass matrix is diagonal, real positive
\ba\label{quarkmass}
m=\left(
\begin{matrix} m_u & &\\ & m_d & \\ &  & m_s\end{matrix}\right).
\ea
In that basis the observable CP-odd parameters are encoded in $d_{ij}$, and of course $\theta$. In the convention used in \eqref{Lag} (we have an imaginary $i$ in front of the dipole), the hermitian part of $d_{ij}$ describes quark EDMs, whereas the anti-hermitian part corresponds to quark magnetic dipoles. Because we will always work in a single insertion approximation, the latter will play no role in our analysis and can be neglected. We can hence assume $d_{ij}$ is hermitian without loss of generality. By charge conservation, the quark dipole must be diagonal in the up-quark, but in general can contain a mixing between the down and the strange:
\begin{align}\label{dgen}
d&=\left(
\begin{matrix} d_u & &\\ & d_d & d_{sd}^*\\ & d_{sd} & d_s\end{matrix}\right).
\end{align} 
The parameter $d_{sd}$ mediates $\Sigma\to N, \Xi\to\Sigma,\Lambda\to N, \Xi\to\Lambda$ transition dipoles (see Appendix \ref{app:chiralPT}) as well as rare $s\to d$ processes. Kaon decays such as $K\to\pi\pi\gamma$ are not very constraining and bounds from $K_0-\overline{K}_0$ mixing likely dominate. The parameter $d_{sd}$ will not be directly relevant to us because in a single insertion approximation it does not affect the neutron EDM, but to be completely general we will use the form \eqref{dgen}.

Because the contribution to the vacuum energy proportional to $d$ must be accompanied by a photon loop, the presence of a QCD axion would affect \eqref{nEDM} first at order $e^2/(4\pi)^2$. Our analysis of the leading order expression \eqref{nEDM} therefore remains unchanged if a QCD axion is introduced (additional comments can be found in Appendix \ref{sec:chromo}).

\subsection{Tensor charges}
\label{sec:tensor}

Throughout the paper we will assume that $F_{\mu\nu}$ reduces to a constant external electric field along the $3$-axis, i.e. $F_{03}=-F_{30}\equiv E_{\text{el}}$ with all other entries vanishing. The neutron EDM is conventionally defined as (minus) the correction to the neutron energy linear in the electric field. At linear order in $E_{\text{el}}$ the QCD Hamiltonian is perturbed by a CP-violating operator which can be read from \eqref{Lag}
\ba\label{Ham}
\Delta H=\int {\text{d}}{\bf x}\,\frac{d_{ij}}{2}\left(\overline{q}_ii\sigma^{\mu\nu}\gamma_5q_j\right)_{\Lambda_{\text{UV}}}F_{\mu\nu}, 
\ea
where the subscript $_{\Lambda_{\text{UV}}}$ reminds us that \eqref{Lag} is defined at that scale. Denoting by $|n\rangle$ the state of a neutron at rest with spin $J_3=+1/2$ along the third axis, we therefore define
\ba\label{EB}
\frac{\langle n|\Delta H|n\rangle}{\langle n|n\rangle}=-d_nE_{\text{el}}.
\ea
Eq. \eqref{EB} translates into \eqref{nEDM} provided we introduce the tensor charges $g_{Tn}^i$, defined as\footnote{We include a subscript $_n$ to distinguish the neutron matrix elements from the proton tensor charges, which are conventionally called $g_T^{u,d,s}$ in some literature.}
\ba\label{matrixelement}
\langle n|(\overline{q}_ii\sigma^{\mu\nu}\gamma_5q_j)_{\Lambda_{\text{UV}}}|n\rangle=\delta_{ij}\,g_{Tn}^j\,\overline{u}i\sigma^{\mu\nu}\gamma_5u,
\ea
where $u$ is the neutron spinor (with spin up) normalized such that $\overline{u}u=2M_n$, with $M_n$ the neutron mass. This expression is exact in the QCD coupling, and first non-trivial order in $d$. There are no contributions to $d_n$ from the off-diagonal $d_{sd}$ because QCD conserves flavor. Note that the matrix element \eqref{matrixelement} is scale-dependent. Unless specified otherwise, in the following by $g_{Tn}^i$ we mean the form factor defined at $\Lambda_{\text{UV}}$.

Several analytical estimates of the tensor charges have been put forward in the past. According to naive dimensional analysis (NDA) one should expect 
\ba\label{NDA}
|g_{Tn}^{u,d,s}|\sim1~~~~~~~({\text{NDA}}), 
\ea
though it is difficult to be more quantitative. The tensor charges of the valence quarks are intuitively of order unity because there is an ${\cal O}(1)$ amount of up and down quarks in the nucleons. The strange contribution arises from closed fermion loops with virtualities set by the QCD scale. At that scale the QCD coupling is large and the strange mass may be considered small, and the effect of the strange loops is presumably unsuppressed as well. The only sense in which those diagrams would be considered ``small" is in a large $N_c$ expansion, in which $|g_{Tn}^{s}|\sim1/N_c$. We will see this more explicitly in Section \ref{sec:largeN}. But, needless to say, even that suppression would not be numerically significant in our world with $N_c=3$. 

As a more quantitative approach we may consider the non-relativistic quark model (NRQM). In that picture one {\emph{naively}} expects 
\ba\label{NRQMgT}
\frac{g_{Tn}^d}{g_{Tn}^u}=-4,~~~~~~~g_{Tn}^s=0~~~~~~~({\text{Naive NRQM}}), 
\ea
where the vanishing of $g_{Tn}^s$ appears as a trivial consequence of the strange quark being a sea quark. Importantly, though, this is truly just a ``naive" expectation. The relevant degrees of freedom in that model have masses set by the dynamical scale of QCD and should therefore be interpreted as constituent quarks. Hence, what the non-relativistic quark model really says is that the dipole of the {\emph{constituent strange}} does not contribute to $d_n$. Thus, it is impossible to make a concrete prediction of the tensor charges $g_{Tn}^{u,d,s}$ in the NRQM without a clear mapping between constituent quark dipoles and $d_{u,d,s}$. Since this important point has apparently been missed in earlier studies, we will discuss it in Section \ref{sec:NRQM}.

Another popular method proposed to estimate $g_{Tn}^{u,d,s}$ relies on QCD sum rules \cite{He:1994gz,Pospelov:2000bw,Hisano:2012sc}. In that language the strange quark contribution appears at a high perturbative order in the OPE. For this reason it has been ignored in the calculations carried out so far, though there is a priori no reason for a perturbative expansion to be of any guidance at the scales of interest. Since only valence quarks have been included, the predictions of the QCD sum rule method obtained until now turn out to agree with \eqref{NRQMgT}.

The only rigorous way to evaluate $g_{Tn}^{u,d,s}$ is performing numerical simulations with lattice QCD techniques. Of course, this path is not all roses either, since the numerical evaluation of $g_{Tn}^{u,d,s}$ requires the precise determination of disconnected diagrams, which are computationally very costly \cite{Aoki:1996pi}, and of the matching between the discretized and the continuum descriptions, on which we will come back in Section \ref{sec:conclusions}. An up to date summary of the current status of lattice QCD investigations is given by the latest report \cite{FlavourLatticeAveragingGroupFLAG:2024oxs} of the Flavor Lattice Averaging Group. The simulations of various collaborations, and in particular the most accurate ones currently carried out by \cite{Bhattacharya:2016zcn,Gupta:2018lvp,Park:2025rxi}, indicate that 
\ba\label{latticeQCD}
\frac{g_{Tn}^d}{g_{Tn}^u}\approx-4,~~~~~~~\frac{|g_{Tn}^s|}{|g_{Tn}^{u,d}|}\approx0.002\div0.01~~~~~~~({\text{Lattice QCD}}).
\ea
This result is surprisingly close to that derived from a naive interpretation of the NRQM. Furthermore, the measured value of $g_{Tn}^s$ departs dramatically from NDA expectations; that in itself is a remarkable fact, since NDA usually works quite well in QCD. If NDA fails so badly, it means that our understanding of the underlying physics is evidently inadequate. We therefore have to understand the origin of the results in \eqref{latticeQCD}, even if just at a qualitative level.

As already emphasized in the introduction, the strong suppression of the strange matrix element in \eqref{latticeQCD} is apparently a peculiarity of simulations of the tensor charges, and is not observed in the numerical evaluation of the strange scalar and axial-vector nucleon matrix elements. Perhaps the reason behind this difference stems from the $C$-odd nature of the tensor operator \cite{Falk:1999tm}. In the evaluation of the matrix element of a C-odd operator $\overline q\Gamma q$ the contribution of the $q$ anti-quark tends to compensate that of the quark $q$, and so the effect of a sea tends to be suppressed compared to that of a valence quark. More explicitly, in the study of polarized deep inelastic scattering, refs. \cite{Jaffe:1991kp,He:1994gz} show that the form factor $g_{Tn}^i(\Lambda)$ evaluated at the scale $\Lambda$ can be viewed as a measure of the number of transversely-polarized partons of type $i$ {\emph{minus}} the number of transversely-polarized anti-partons of type $i$ at that scale. This argument implies that sea partons, like the strange, should have a vanishing tensor charge. The crucial relative minus sign of the parton and anti-parton contribution, originating from the C-odd nature of the tensor operator, distinguishes the tensor charges from the matrix elements of scalars and axial-vector currents, the latter being\footnote{Ref. \cite{Ellis:1996dg} identifies $g_{Tn}^{i}$ with the matrix elements $g_{An}^i$ of the quark axial currents. However that identification only applies in the limit of non-relativistic quarks and is not expected to be accurate.}
\ba\label{matrixelementA}
\langle n|(\overline{q}_i\gamma^\mu\gamma_5q_j)_{\Lambda_{\text{UV}}}|n\rangle=\delta_{ij}\,g_{An}^j\,\overline{u}\gamma^\mu\gamma_5u.
\ea
It should however be noticed that, while suggestive, this observation cannot explain the small $g_{Tn}^s$ quoted in \eqref{latticeQCD}. The conclusion that $g_{Tn}^s(\Lambda)$ vanishes relies entirely on the parton interpretation, which does not provide an accurate description of QCD at the scales $\Lambda\sim\Lambda_{\text{UV}}$ we are interested in. The reason is that the parton picture, by definition, ignores corrections controlled by powers of the non-perturbative scale and the quark masses. In Subsection \ref{sec:RG} we will see that $g_{Tn}^s$ is in fact {\emph{entirely}} due to IR-sensitive corrections proportional to powers of $M_h/\Lambda$ and $m/\Lambda$. As a result, the C-odd nature of the tensor does not imply $|g_{Tn}^s|$ must be suppressed. 

\subsection{A look at the diagrams}
\label{sec:RG}

The Feynman diagrams that contribute to $g_{Tn}^s$ are saturated at the QCD scale and are therefore incalculable. This is qualitatively different from the contribution of a quark heavier than the QCD scale, whose effect is controlled by diagrams with a virtuality of order the heavy quark mass and therefore within the perturbative regime.

\begin{figure}[t]
\begin{center}
\includegraphics[width=.27\textwidth]{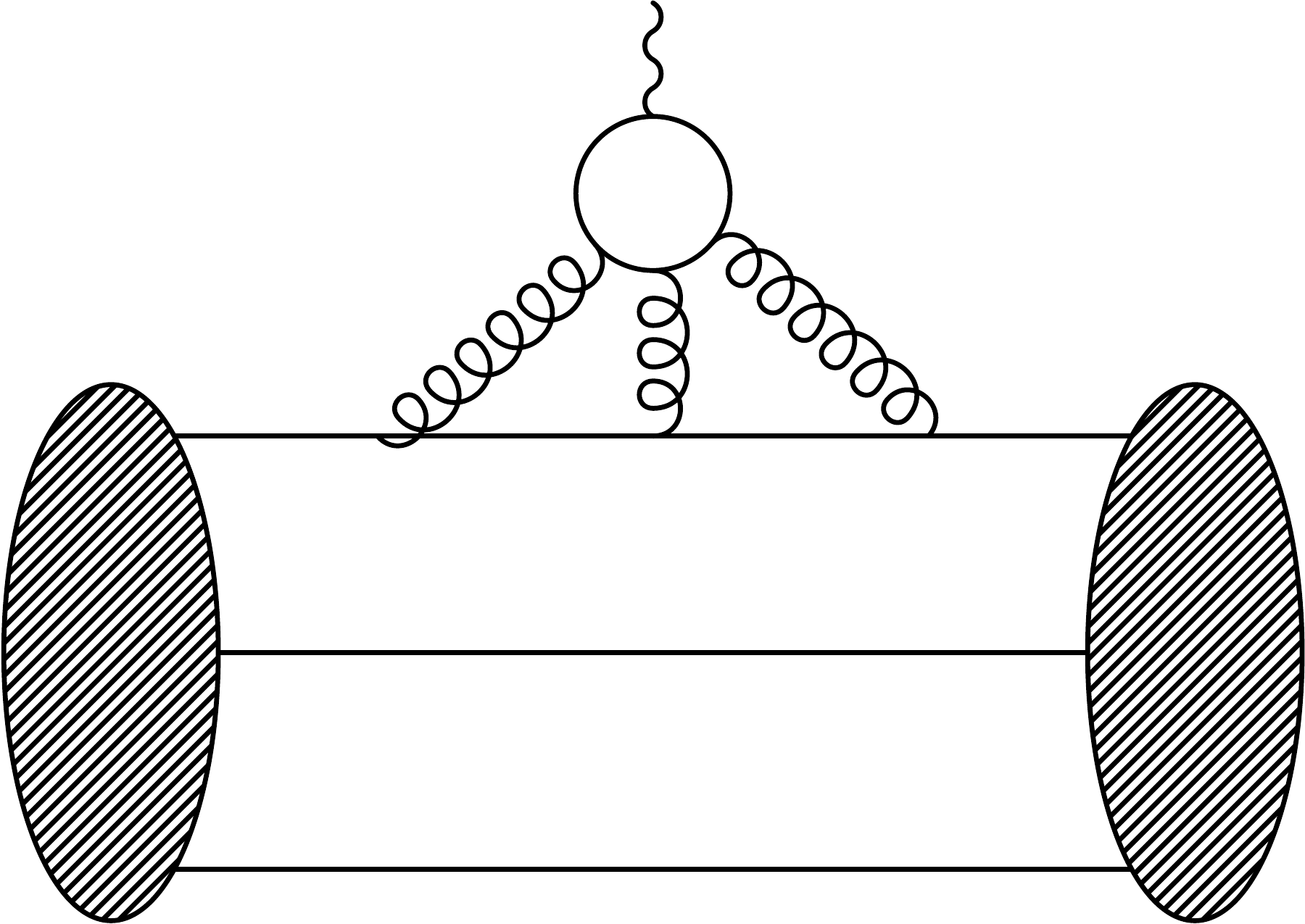}
~~~~~~~~~
\includegraphics[width=.27\textwidth]{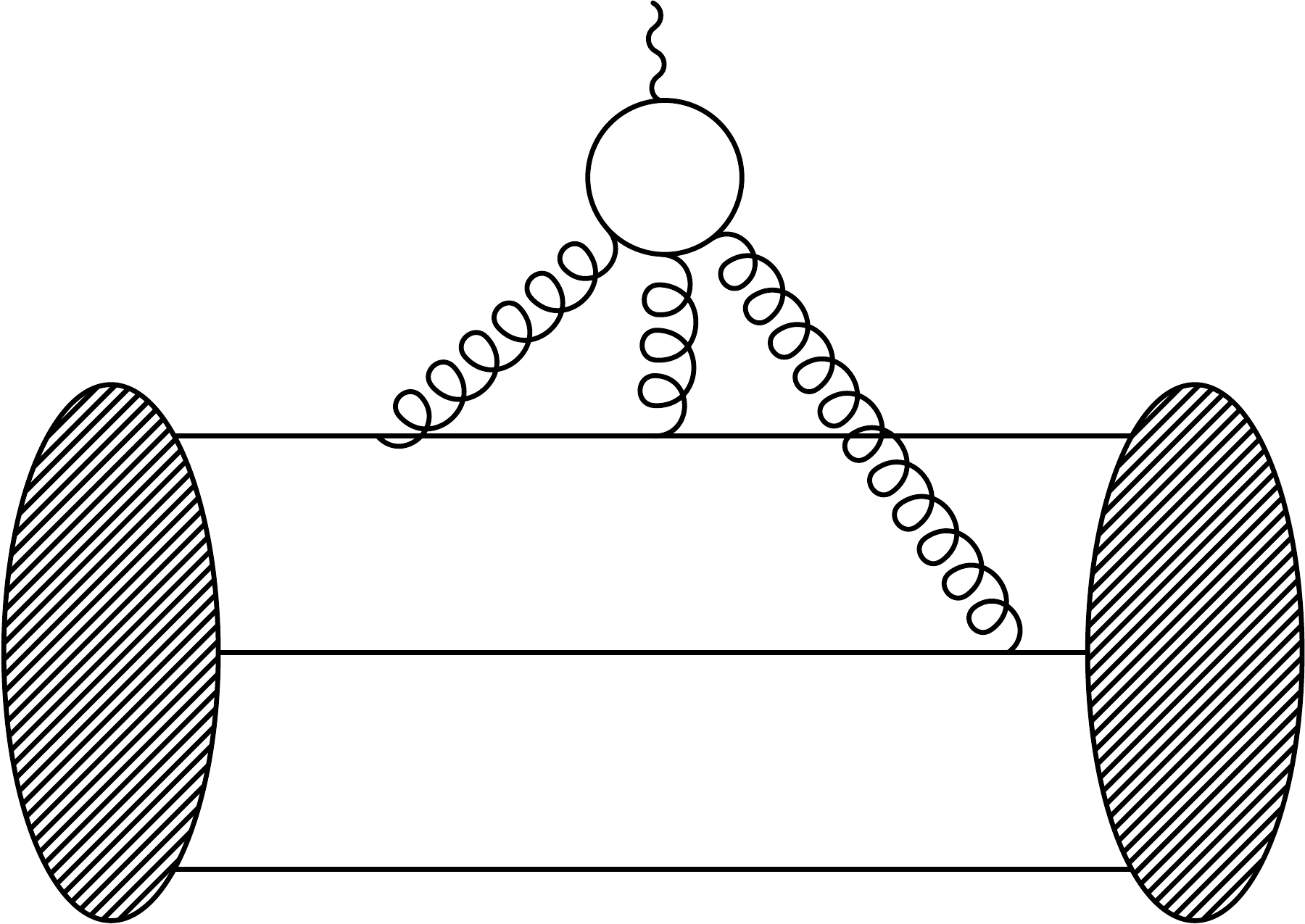}
~~~~~~~~~
\includegraphics[width=.27\textwidth]{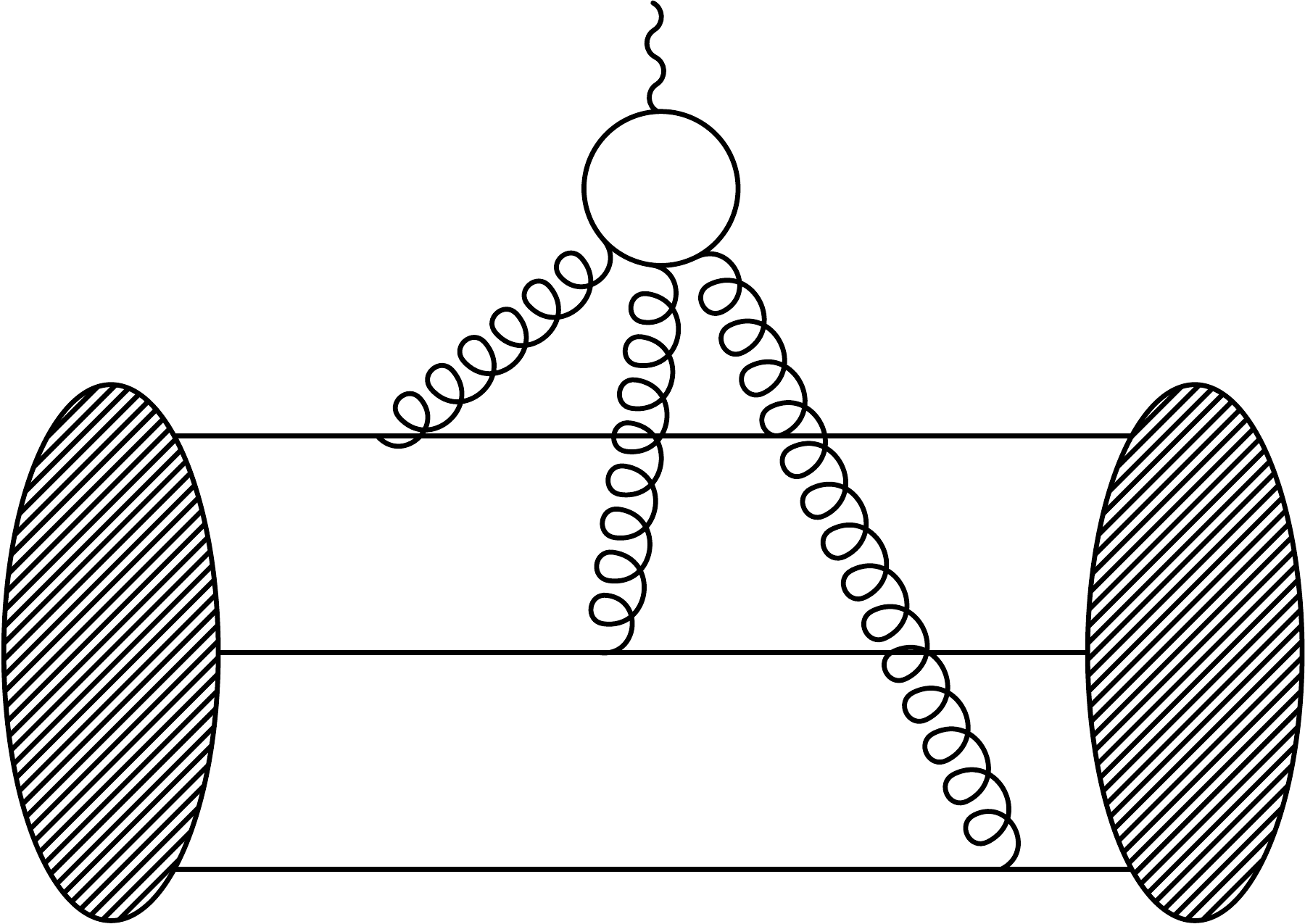}
\caption{\small Examples of the first class of diagrams contributing to $g_{Tn}^s$. The wavy line represents the external electric field and is attached to a closed strange loop. The horizontal solid lines are valence quarks and the curly lines are virtual gluons.}\label{FigDI}
\end{center}
\end{figure}

The contributions to $g_{Tn}^s$ can be divided into two classes. In the first class we find diagrams with a closed strange loop connected to the valence quarks by an appropriate number of gluons. With no momentum flowing in the tensor operator, it is easy to see that Parity combined with Bose statistics force the number of gluon legs to be at least three. An example of a few representative diagrams of this type is shown in Fig. \ref{FigDI}. Clearly that topology includes analogous diagrams with a loop of up and down quarks replacing that of the strange. By chiral invariance, the coupling $d_{ij}$ must appear in a flavor trace multiplied by a positive power of a quark mass from a quark propagator, the simplest option being ${\text{Tr}}[dm^\dagger]$. A mere dimensional argument is then sufficient to convince oneself that the resulting contribution to $g_{Tn}^i$ will have to contain inverse powers of the quark mass, namely the only available scale in the calculation. Because here we are interested in the light quarks $u,d,s$, this unavoidably implies that the loop is dominated by scales at which QCD is non-perturbative.\footnote{On the other hand, if we were analyzing the effect of dipoles of heavy quarks, a perturbative analysis like that of \cite{Ema:2022pmo} would suffice. In that case the first diagram in Fig. \ref{FigDI} would be mapped into a light quark dipole with coefficient $\propto m{\text{Tr}}[d_{\text{heavy}}m_{\text{heavy}}^\dagger]/|m_{\text{heavy}}|^2$ times a loop factor, whereas the other diagrams would correspond to higher-dimensional operators involving the light quarks and suppressed by powers of $m_{\text{heavy}}$.}

In a theory with spontaneous symmetry breaking one should also take into account diagrams with insertions of the order parameter. In the present case these consist of amplitudes with additional pairs of external quark legs compared to Fig. \ref{FigDI}. Examples of this class of contributions to $g_{Tn}^s$ may be obtained by cutting open the quark loop in Fig. \ref{FigDI} and adding appropriate gluon lines. After chiral symmetry breaking, the additional external legs combine to form a quark condensate and the resulting contributions can display flavor structures, among others, of the form ${\text{Tr}}[d\langle\overline qP_Rq\rangle^\dagger]$, which are parametrically larger than the ${\text{Tr}}[d m^\dagger]$ mentioned earlier. Yet, the diagrams of this second class are obviously IR-sensitive as well, and therefore incalculable.\footnote{If we were considering the effect of heavy quark dipoles, the contributions of this second class would be parametrized by higher dimensional operators with several light quarks and suppressed by powers of $m_{\text{heavy}}$.}

To quantitatively estimate $g_{Tn}^s$ one is therefore forced to introduce an IR-regulator at $\Lambda>M_h$ on the loop virtualities. This can be achieved for example inserting a momentum scale of order $\Lambda$ on the external states. Following this prescription one would find that the two classes of diagrams mentioned above induce corrections to $g_{Tn}^s$ that scale with powers of $m/\Lambda$ and $M_h/\Lambda$, as anticipated around \eqref{matrixelementA}. 

Alternatively, one can approach the problem adopting a Wilsonian effective action defined at scales $\Lambda$ satisfying $M_h<\Lambda<\Lambda_{\text{UV}}$. In that language our diagrams are mapped into operators with arbitrary pairs of quark fields. For example, the diagrams in Fig. \ref{FigDI} are parametrized respectively by operators with two, four, six external quarks plus quark mass insertions. Diagrams of the second class have more fermionic fields and are commonly known as ``dangerous irrelevant operators". Among them we find for example the following
\begin{align}\label{LagW}
{\cal O}_\Lambda=[d_{mn}\overline{q}_mP_Rq_n][\overline{q}_jP_Lq_i][\overline{q}_ii\sigma^{\mu\nu} P_Rq_j]\,F_{\mu\nu}+{\text{hc}},
\end{align}
where $[\cdots]$ denotes contraction of the color and spinor indices. This has a coefficient of order\footnote{We will see how these expressions scale at large $N_c$ in Section \ref{sec:largeN}. For now we work at finite $N_c$.}
\ba
C(\Lambda)\sim\frac{g^4}{\Lambda^6}\left(\frac{g^2}{16\pi^2}\right)^n,
\ea
for some loop order $n$. The operator \eqref{LagW} has the correct quantum numbers to mix with \eqref{Ham} along the {\emph{Wilsonian}} RG evolution down to $\Lambda$, and hence to contribute to the neutron EDM. Obviously, for $\Lambda\gg M_h$ its effect is suppressed by a small loop factor and inverse powers of $\Lambda$. However, as we approach the hadron scale it might become numerically significant. According to NDA the QCD coupling gets maximally strong at that scale, $g^2\to16\pi^2$, and $C({M_h})\sim(16\pi^2)^2/M_h^6$ becomes of natural size. After chiral symmetry breaking, the condensate $\langle\overline q_iq_j\rangle\sim\delta_{ij}M_h^3/(16\pi^2)$ forms, and the effective action at the hadronic scale includes
\begin{align}\label{LagW1}
{\cal L}_{M_h}\supset C(M_h){\cal O}_{M_h}
&\supset{\cal O}(1) [d_{mn}\delta_{mn}]\,[\overline{q}_ii\sigma^{\mu\nu} \gamma_5q_i]\,F_{\mu\nu}.
\end{align}
Interestingly, the latter describes a flavor-diagonal dipole operator proportional to ${\text{Tr}}[d]$, which of course contains the strange EDM as well. In the Wilson language we thus find a sizable mixing between the valence and the sea tensors. In the alternative prescription with an IR-momentum flowing in the diagrams, these would correspond to sizable finite threshold effects.

Diagrams with additional powers of the order parameter are suppressed by further powers of $g^2/\Lambda^3$, but when approaching $\Lambda\to M_h$ they should all be re-summed as they are expected to contribute comparably to \eqref{LagW1}.

In our Wilsonian EFT there are also operators with insertions of both the quark condensate and quark masses. For example, we have some analogous to \eqref{LagW} with the quark bilinears replaced by powers of $m$, 
\begin{align}\label{LagW2}
{\cal L}_{\Lambda}\supset\frac{g^2}{\Lambda^4}\left(\frac{g^2}{16\pi^2}\right)^{n'}[d_{mn}m^*_{mn}][\overline{q}_jP_Lq_i][\overline{q}_ii\sigma^{\mu\nu} P_Rq_j]\,F_{\mu\nu}+{\text{hc}},
\end{align}
and
\begin{align}\label{LagW3}
{\cal L}_{\Lambda}\supset\frac{1}{\Lambda^2}\left(\frac{g^2}{16\pi^2}\right)^{n''}[d_{mn}m^*_{mn}][m_{ij}][\overline{q}_ii\sigma^{\mu\nu} P_Rq_j]\,F_{\mu\nu}+{\text{hc}}.
\end{align}
As we approach the hadron scale, and again using NDA, these induce corrections to \eqref{LagW1} of order $m/M_h$ and $m^2/M_h^2$, respectively.

There are also operators of lower engineering dimension than shown so far. Performing a one-instanton calculation for example we find, up to an order one factor, 
\begin{align}\label{instanton}
{\cal L}_{\Lambda}\supset\frac{1}{\Lambda}e^{-{8\pi^2}/{g^2}-i\theta}\left(\frac{16\pi^2}{g^2}\right)^6\epsilon_{ii'i''}\epsilon_{jj'j''}d^*_{i'j'}{m}^*_{i''j''}\,[\overline{q}_ii\sigma^{\mu\nu} P_Rq_j]\,F_{\mu\nu}+{\text{hc}},
\end{align}
where
\ba
\epsilon_{ii'i''}\epsilon_{jj'j''}d^*_{i'j'}{m}^*_{i''j''}=
\left(\begin{matrix}
d_d^*m_s^*+d_s^*m_d^* & & \\
 & d_u^*m_s^*+d_s^*m_u^* & - d_{sd}^*m_u^*\\
 & -d_{ds}^*m_u^* & d_u^*m_d^*+d_d^*m_u^*
 \end{matrix}\right).
\ea
At around the QCD scale instantons become unreliable and the best we can do is, again, to invoke NDA and replace $\exp(-{8\pi^2}/{g^2})/\Lambda$ with $1/M_h$. Effects similar to these are at the origin of $[m^\dagger]^{-1}{\text{Det}}[m]$ corrections in chiral perturbation theory (related to the so-called Kaplan-Manohar ambiguity \cite{Kaplan:1986ru}). In the present case, the main implication of \eqref{instanton} would be a mixing between $g_{Tn}^u$ and $g_{Tn}^d$, potentially of order $m_s/M_h=20\%$. On the other hand, $g_{Tn}^s$ would only receive negligible corrections proportional to the light quark masses.

To summarize, the diagrams contributing to $g_{Tn}^s$ are either proportional to powers of $m$ or the condensate $\langle \overline qq\rangle$, or both, divided by some IR scale $\Lambda$. It is impossible to reliably calculate their impact at the hadronic scale analytically and so we had to resort to NDA, obtaining essentially a refinement of our earlier guess \eqref{NDA}. But we have reasons to expect that our naive estimates should provide a reasonable order one approximation of the actual result. The Wilson operators proportional to powers of $m$, as in eqs. \eqref{LagW2}, \eqref{LagW3}, and \eqref{instanton}, are the progenitors of the {\emph{counterterms}} encountered in a chiral expansion in $m/M_h$, derived by matching QCD to a theory of baryons and mesons. They are the counterparts of the genuinely IR contributions, non-analytic in $m$, arising from chiral loops, i.e. from integrating down to $\Lambda\ll M_h$. The latter are evaluated in Section \ref{sec:chiralPT} at first non-trivial order in $m/M_h$, and found to be numerically of the order suggested by NDA. Barring unnatural cancellations, then, the same must be true for the local counterterms, namely for the coefficients of operators like \eqref{LagW2}\eqref{instanton} evolved down to $\Lambda\sim M_h$ and, reasonably, also for ${\cal O}(m^2)$ counterterms like \eqref{LagW3} and the parametrically larger contributions like \eqref{LagW1}. Our exercise in chiral perturbation theory (see Section \ref{sec:chiralPT}) can therefore be interpreted as an indirect confirmation of the NDA estimates of this section.

\section{Large $N_c$ counting}
\label{sec:largeN}

The large $N_c$ expansion is the only systematically improvable approximation of a continuum non-abelian gauge theory at strong coupling. Since in the real world $N_c=3$, one may suspect that a perturbative expansion in powers of $1/3$ cannot converge fast enough to make concrete predictions. Yet, there is plenty of evidence that large $N_c$ techniques provide a qualitatively accurate description in many instances. We hope the same applies to our question.

The analysis of baryons at large $N_c$ was initiated in \cite{Witten:1979kh} and then subsequently developed by various authors, see in particular \cite{Dashen:1993jt,Dashen:1994qi,Luty:1993fu,Luty:1994ua}. The crucial difference between the large $N_c$ counting for mesons and that of baryons is that the latter are composed of $N_c$ quarks, and therefore the multiplicity of diagrams plays a crucial role. Another important difference compared to mesons is that with three light flavors there is some ambiguity in the extrapolation of ordinary baryons to large $N_c$. Specifically, the flavor $SU(3)_V$ multiplet of spin $J=1/2$ has components with strangeness $S=0,-1,-2,\cdots,-(N_c+1)/2+1,-(N_c+1)/2$, where we assumed $N_c$ is odd. It is therefore not obvious to which representation the strange baryons in the real world should be associate to since, for example, the $\Sigma$ may be identified with either the component with $S=-1$ or the one with $S=-(N_c+1)/2+1$. The strangeness is the same at $N_c=3$ but clearly is not at arbitrary $N_c$, and the predictions of a large $N_c$ analysis in those two cases can differ qualitatively.

Fortunately, this ambiguity has no implications when analyzing the physics of the nucleons at leading $1/N_c$ order. The leading order predictions are in fact the same as long as we interpret the nucleon as an isospin doublet with strangeness $|S|\ll{\cal O}(N_c)$. To appreciate that, consider first $g_{Tn}^u$. Since our neutron is an isospin doublet with ``small to vanishing strangeness", it must be a state composed of ${\cal O}(N_c)$ up and down quarks and a number $\ll{\cal O}(N_c)$ of strange quarks. The dipole $d_u$ can then be attached at tree level on any of the valence up quarks. As a result we have $g_{Tn}^u={\cal O}(N_c)$. Loops do not alter this conclusion. Similar considerations show that $g_{Tn}^d={\cal O}(N_c)$. For what concerns $g_{Tn}^s$, instead, tree effects are negligible, or absent altogether if $S=0$. We should therefore go back to the loops of Fig. \ref{FigDI} and start, say, by estimating the diagram on the left, where a strange loop is attached to the same valence quark via the gluon lines. The fermion loop brings a suppression ${\cal O}(1/N_c)$, but because there is a number ${\cal O}(N_c)$ of lines on which to attach the three gluons the overall result is $g_{Tn}^s={\cal O}(1)$. Consistently, the same conclusion extends to the other kinds of diagrams. For example, for diagrams in which two gluon legs are attached to the same valence quark and the third to another one (see the middle diagram in Fig. \ref{FigDI}), one can verify that the loop is suppressed by ${\cal O}(1/N_c^2)$. Yet, there are ${\cal O}(N^2_c)$ ways to select the two valence lines on which the gluons are attached. Therefore, the net result is again $g_{Tn}^s={\cal O}(1)$. The same conclusion holds for the last diagram in Fig. \ref{FigDI}, whereas one can show that $g_{Tn}^s\leq{\cal O}(1)$ from diagrams with more gluon legs. All in all, large $N_c$ considerations lead to $g_{Tn}^{u,d}={\cal O}(N_c)$ and $g_{Tn}^s={\cal O}(1)$.

We can say more if we postulate that flavor $SU(3)_V$ is an accurate symmetry of nature. When framed in that context, the statement that fermionic loops are suppressed becomes (see Eq. \eqref{LargeN-LO} and Eq. \eqref{holds} below for a discussion of flavor $SU(3)_V$)\footnote{Analogously to \eqref{SU3relation}, combining large $N_c$ and flavor $SU(3)_V$ the quark contributions to the proton spin satisfy $g_{A}^u+g_{A}^d+g_{A}^s=0$ (see the explicit definition in \eqref{matrixelementA}). The latter relation was approximately satisfied in several experimental analyses of inclusive deep inelastic scattering, where $SU(3)_V$ was in fact imposed (see, e.g. \cite{Brodsky:1988ip}).}
\ba\label{SU3relation}
g_{Tn}^u+g_{Tn}^d+g_{Tn}^s={\cal O}(1).
\ea
This relation, when combined with the earlier result $g_{Tn}^s/g_{Tn}^{u,d}={\cal O}(1/N_c)$ gives
\ba\label{diagrammN}
\frac{g_{Tn}^d}{g_{Tn}^u}=-1+{\cal O}(1/N_c),~~~~~~~\frac{g_{Tn}^s}{g_{Tn}^u}={\cal O}(1/N_c)~~~~~~~(SU(3)_V{\text{-Large $N_c$}}).
\ea
Eq. \eqref{diagrammN} should be added to the list of estimates collected in Section \ref{sec:tensor}.\footnote{A discussion of the tensor charges for the proton at large $N_c$ is found in \cite{Olness:1992zb}.}

Famously, the $SU(3)_V$ flavor symmetry is explicitly broken by the quark masses and electromagnetism. The dominant effect is by far due to the strange quark mass and is $\sim m_s/M_h$, which is numerically comparable to $1/N_c$. The latter modifies $g_{Tn}^i$ via calculable non-analytic corrections (starting at order $m_s\ln m_s$) as well as incalculable power-law contributions. The leading non-analytic corrections in $m_s/M_h$ will be carefully analyzed in Section \ref{sec:chiralPT}. Here we just mention that the leading order counterterms induce corrections of order 
\ba\label{diagrammN1}
\frac{\delta g_{Tn}^s}{g_{Tn}^u}={\cal O}\left(\frac{1}{N_c}\frac{m_s}{M_h}\right).
\ea 

These considerations represent important consistency checks for any quantitative analysis of the tensor charges. For example, they imply that at large $N_c$ all contributions discussed in Section \ref{sec:RG} are at least of order $1/N_c$ (the one in \eqref{instanton} is actually down by $e^{-{\cal O}(N_c)}$ in this limit). We will see that the more quantitative results derived below are indeed consistent with large $N_c$ expectations.

In the following subsection we will confirm the large $N_c$ scaling via the more systematic techniques introduced in \cite{Dashen:1993jt,Dashen:1994qi,Luty:1993fu,Luty:1994ua}.

\subsection{A systematic approach}

By carrying out a diagrammatic expansion on a reference ``baryon ground state", refs. \cite{Dashen:1994qi,Luty:1993fu} argued that the matrix elements of static baryons can be written as matrix elements of ``many-body" interactions involving quarks of definite spin and $SU(3)_V$ flavor.\footnote{As usual, we can write our results in a $SU(3)_L\times SU(3)_R$ invariant way introducing the Nambu-Goldstone matrix $\xi$ (see Section \ref{sec:NRQM}) and replacing $d$ with combinations of $\frac12 \xi^\dagger d\xi^\dagger$ everywhere in this section.} Concretely, the matrix element of the dipole operator with a static baryon state is written, at all orders in QCD and leading order in $d$, as 
\begin{align}\label{LutyME}
d_{BB'}=(B'|{\cal O}_3|B).
\end{align}
The object $|B)={\cal B}^{i_1\alpha_1\cdots i_{N_c}\alpha_{N_c}}\alpha_{i_1\alpha_1}^\dagger\cdots\alpha_{i_{N_c}\alpha_{N_c}}^\dagger|0)$ denotes a properly normalized baryon state of spin and flavor determined by the wavefunction ${\cal B}$. Since the color Levi-Civita tensor has been absorbed in ${\cal B}$, $\alpha^{i\alpha}/\alpha_{i\alpha}^\dagger$ are {\emph{commuting}} annihilation/creation operators for quarks of spin $\alpha=+,-$ and flavor index $i$. The quantity ${\cal O}_k$ in \eqref{LutyME} is a linear combination of the following (hermitian) operators
\ba\label{quant}
\left\{d J_k\right\},~~~{\text{Tr}}[d]\,\left\{J_k\right\},~~~\left\{d\right\}\left\{J_k\right\},~~~\left\{d J_{k'}\right\}\left\{J_{k'}\right\}\left\{J_k\right\},
\ea
multiplied by arbitrary powers of $\left\{J_{k'}\right\}\left\{J_{k'}\right\}$, where $J_k=\sigma_k$ is the spin operator \cite{Dashen:1994qi,Luty:1993fu}. We adopt the notation \cite{Luty:1993fu,Luty:1994ua}
\ba
\left\{W J_k\right\}=W^i_{j}\alpha_{i\alpha}^\dagger[J_k]^\alpha_\beta\alpha^{j\beta}.
\ea
Structures of the form $\left\{\cdots\right\}$ are referred to as $1$-body operators, and consequently $\left\{\cdots\right\}^n$ as $n$-body.

Neglecting the $SU(3)_V$ breaking due to the quark masses and electromagnetism, the most important contributions to the baryon EDM are given by
\begin{align}\label{OgenA}
{\cal O}_k
&=c_0\left\{d_a\lambda_a J_k\right\}+\frac{c_1}{N_c}\,{\text{Tr}}[d]\left\{J_k\right\}+\frac{c_2}{N_c}\left\{d_a\lambda_a\right\}\left\{J_k\right\}+\frac{c_3}{N_c^2}\left\{d_a\lambda_aJ_{k'}\right\}\left\{J_{k'}\right\}\left\{J_k\right\}+\cdots,
\end{align}
where we separated the trace from the traceless parts of the dipole as 
\ba
d=\frac13{\text{Tr}}[d]+d_a\lambda_a, 
\ea
with $\lambda_a$ the Gell-Mann matrices. The coefficients $c_{0,1,2,3}$ are of order unity plus corrections proportional to increasing powers of $1/N_c$. To understand the inverse powers of $N_c$ in \eqref{OgenA} one observes that r-body operators have matrix elements scaling at most as $N_c^r$ (for baryons of any spin). Yet, the energy shift due to an ${\cal O}(1)$ external electric field can be at most of order the baryon mass, namely ${\cal O}(N_c)$. The very hypothesis that an expansion exists therefore indicates the coefficients $c_{0,1,\cdots}$ must satisfy upper bounds, which we assume get saturated in the spirit of NDA. This logic readily explains the size of $c_{0,2,3}$. The power of $1/N_c$ in front of $c_1$ is understood noting that the trace of $d$ can only occur because of a fermionic loop, and the latter are $1/N_c$-suppressed.

We now want to evaluate the matrix element \eqref{LutyME} of \eqref{OgenA} for ${\cal B}$ being the appropriate neutron wavefunction. Conservatively assuming the neutron is associated to a state with vanishing strangeness ($S=0$), it is easy to see that the neutron corresponds to \cite{Dashen:1993jt,Dashen:1994qi,Luty:1994ua}
\ba\label{normn}
|n)=C_n\alpha_{d+}^\dagger(\alpha_{u+}^\dagger\alpha_{d-}^\dagger-\alpha_{u-}^\dagger\alpha_{d+}^\dagger)^{\frac{N_c+1}{2}}|0),
\ea 
with $C_n$ a normalization chosen so that $(n|n)=1$. Using this definition we find that the matrix element of $\left\{\lambda_aJ\right\}$ can be of order $N_c$ (see Appendix \ref{app:onebody}). On the other hand, all other contributions in \eqref{OgenA} are at most ${\cal O}(1)$. 
Hence, the largest contribution to the matrix element \eqref{LutyME} is from the $c_0$ term in \eqref{OgenA}. We find (see Appendix \ref{app:onebody})  
\begin{align}\label{LargeN-LO}
\left.d_n\right|_{\text{LO}}
&=c_0\,(n|\left\{d_a\lambda_aJ_3\right\}|n)\\\no
&=c_0\left[\left(-\frac{N_c+1}{6}\right)d_u+\left(\frac{N_c+3}{6}\right)d_d+\left(-\frac{1}{3}\right)d_s\right].
\end{align}
Each of the three coefficients $g_{Tn}^{u,d,s}$ is also affected by independent ${\cal O}(1)$ corrections due to the operators in \eqref{OgenA}. The explicit result \eqref{LargeN-LO} satisfies \eqref{SU3relation}, as anticipated, and reduces to \eqref{diagrammN} for $N_c\gg1$. Furthermore, when the effect of the strange quark mass is taken into account, the present framework confirms that additional corrections of the parametric form \eqref{diagrammN1} arise. For example, a matrix element like that shown in \eqref{LargeN-LO} with $d$ replaced by a product of $d$ and $m/M_h$ implies a correction of order $\delta g_{Tn}^s/g_{Tn}^u={\cal O}(\frac{1}{N_c}m_s/M_h)$, as it should be.

In the real world with $N_c=3$ the ${\cal O}(1)$ corrections neglected in \eqref{LargeN-LO} are likely to be too large to be ignored. Yet, one may be tempted to evaluate \eqref{LargeN-LO} at $N_c=3$. That would give 
\ba\label{largeNest}
\frac{g_{Tn}^d}{g_{Tn}^u}=-\frac32,~~~~~~~\frac{g_{Tn}^s}{g_{Tn}^u}=\frac12. 
\ea
A compatible result is obtained via a soliton model of baryons (see Appendix \ref{app:soliton}). In either case, the sizable strange contribution found for $N_c=3$ is probably an overestimate; still, without knowing the relative size of the other coefficients in \eqref{OgenA} nobody can tell by how much.

It is worth stressing that in the present language the prediction of a naive NRQM corresponds to evaluating at $N_c=3$ a matrix element analogous to \eqref{LargeN-LO}, but with the entire $d$ replacing the traceless $d_a\lambda_a$:
\ba\label{LargeN-NRQM}
\left.d_n\right|_{\text{NRQM}}=\left.c(n|\left\{d J_3\right\}|n)\right|_{N_c=3}
=c\left[\left(-\frac{N_c-1}{6}\right)d_u+\left(\frac{N_c+5}{6}\right)d_d\right]_{N_c=3}.
\ea
Comparing \eqref{LargeN-LO} to \eqref{LargeN-NRQM} it becomes apparent that the vanishing of $g_{Tn}^s$ in a naive NRQM occurs because of an exact cancellation between the traceless and the trace parts of $d$. However, there is no underlying symmetry that guarantees the stability of such cancellation, and in fact large $N_c$ treats the traceless and trace parts independently, with the latter being subleading. Analogously, $d_a\lambda_a$ and ${\text{Tr}}[d]$ may appear in different combinations in the discussion of Section \ref{sec:RG}. Consistently, we will show in the next section that ``radiative corrections" within the NRQM tend to split the contributions proportional to the traceless and the trace parts, confirming the expected result $g_{Tn}^s={\cal O}(1/N_c)$.

\section{Constituent quark model}
\label{sec:NRQM}

In this section we critically review the predictions of the non-relativistic quark model (NRQM) and validate the claims of Sections \ref{sec:RG} and \ref{sec:largeN}.

In the NRQM nucleons are described by bound states of weakly-interacting constituent quarks $\psi$ of mass $m_\psi$. This picture provided a surprisingly accurate description of hadronic physics, and it was of course the model that people referred to when the first attempts at estimating the quark dipole contribution to the neutron EDM were made.

In this language the normalized wavefunction of a neutron with spin up is given by 
\begin{align}\label{wavefunN}
|n;{\text{NR}}\rangle=
\frac{1}{3\sqrt{2}}&\left[|ddu\rangle\left(2|\!++\,-\rangle-|\!+-\,+\rangle-|\!-+\,+\rangle\right)\right.\\\no
+\frac{}{}&\left.|dud\rangle\left(2|\!+-\,+\rangle-|\!-+\,+\rangle-|\!++\,-\rangle\right)\right.\\\no
+\frac{}{}&\left.|udd\rangle\left(2|\!-+\,+\rangle-|\!++\,-\rangle-|\!+-\,+\rangle\right)\right],
\end{align}
where $|ddu\rangle$ indicates the flavor content and $|\!++\,\,-\rangle$ the spin of the corresponding flavor. Denoting by $d_\psi$ the EDM matrix of the constituent quarks, the neutron dipole is formally obtained from the tensor product of an operator in flavor space (the constituent quark dipole $d_\psi$) and in spin space (the third component of spin). The result is well-known:
\begin{align}\label{NRQMtree}
\left.d_n\right|_{\text{NRQM}}=\langle n;{\text{NR}}|d_\psi\otimes{\sigma_3}|n;{\text{NR}}\rangle
=-\frac13d_\psi^u+\frac43d_\psi^d.
\end{align}
As already anticipated in the introduction and Section \ref{sec:tensor}, there is no contribution from the constituent strange dipole because the strange is a see quark. The constituent dipoles $d_\psi$ are conceptually different from the bare ones $d$ defined in \eqref{Lag}. The precise relation between the two should be determined by matching the constituent quark picture to QCD at the scale $M_h$.

In order to investigate the relation between $d_\psi$ and $d$ we need to define an unambiguous prescription for the NRQM. The prescription we adopt is the following: the chiral quark model \cite{Manohar:1983md} of Georgi and Manohar is postulated to be the effective field theory description of the constituent quarks; from it the neutron properties are extracted by taking matrix elements of constituent quark operators (renormalized at $M_h$) with the non-relativistic wavefunction \eqref{wavefunN}, analogously to \eqref{NRQMtree}. Needless to say, this is just one of the many prescriptions that reproduce the same predictions of the non-relativistic quark model at leading order in some approximation. The arbitrariness in the choice of the model represents an intrinsic systematic uncertainty of the constituent quark approach.

In the chiral quark model, quarks have a flavor-singlet constituent mass $m_\psi$, interact weakly with gluons and with light mesons compatibly with the full, non-linearly realized, $SU(3)_L\times SU(3)_R$ chiral symmetry. The EFT is defined below the chiral symmetry breaking scale 
\ba\label{Mchi}
M_\chi\equiv\frac{4\pi f_\pi}{\sqrt{N_c}},
\ea
which is numerically of order $M_h$, and reads \cite{Manohar:1983md}: 
\begin{align}\label{chiQM}
{\cal L}_{\text{$\chi$QM}}
&=-\frac{1}{4}G^A_{\mu\nu}G^{A\,\mu\nu}+\overline{\psi}i\gamma^\mu\nabla_\mu\psi-m_\psi\overline{\psi}\psi
\\\no
&+\frac{f_\pi^2}{4}{\text{Tr}}[\partial_\mu U\partial^\mu U^\dagger]+bM_\chi f_\pi^2{\text{Tr}}[{m} U^\dagger+{\text{hc}}]+g_A\overline{\psi}\gamma^\mu\gamma_5A_\mu\psi\\\no
&+\cdots,
\end{align}
with constituent quark fields $\psi$ transforming in the fundamental of color as well as flavor $SU(3)_{V}$, $\nabla_\mu\psi=(D_\mu+V_\mu)\psi$ is the color- and flavor-covariant derivative, and
\begin{align}\label{VA}
V_\mu&=\frac{1}{2}\left(\xi\partial_\mu\xi^\dagger+\xi^\dagger\partial_\mu\xi\right)
\\\no
A_\mu&=\frac{i}{2}\left(\xi\partial_\mu\xi^\dagger-\xi^\dagger\partial_\mu\xi\right).
\end{align} 
The special unitary pion matrix is $\xi=e^{i\Pi}$, with $\Pi=\Pi_aT_a$ and $T_a=\lambda_a/2$ the flavor $SU(3)$ generators normalized such that ${\text{Tr}}[T_aT_b]=\delta_{ab}/2$, and as usual $ U=\xi^2$. The dots in \eqref{chiQM} indicate higher dimensional operators as well as other insertions of $m$. The parameters of the EFT \eqref{chiQM} are formally found by matching the EFT to QCD at the scale $M_\chi$. In our notation $m$ denotes the quark mass renormalized at $\Lambda_{\text{UV}}$ appearing in \eqref{Lag} whereas the parameter $b$ in \eqref{chiQM} is a constant of order unity that includes the RG evolution down to the matching scale.

Taking $m_\psi\sim360$ MeV, $g_A\sim0.75$, and a rather weak gauge coupling, $g^2/(4\pi)\sim0.3$, the Lagrangian \eqref{chiQM} reproduces the predictions of the non-relativistic quark model as well as many other salient features of hadronic physics \cite{Manohar:1983md}. We are thus justified to expect that the same model provides a rough estimate of the neutron EDM as well.

Among the higher dimensional operators included in \eqref{chiQM} there are some that are proportional to the UV coupling $d$ defined in \eqref{Lag}. At leading order in $d$ and zeroth order in the electric charge and ${m}$, the $SU(3)_L\times SU(3)_R$-symmetric operators with the smallest number of derivatives are:
\begin{align}\label{dipNRQM}
{\cal L}_{\text{$\chi$QM}}
\supset-\frac{i}{2}\,\overline{\psi}\hat d_\psi \sigma^{\mu\nu}\gamma_5\psi\,F_{\mu\nu}-\frac{1}{2}\,\overline{\psi}\hat d_\psi' \sigma^{\mu\nu}\psi\,F_{\mu\nu},
\end{align}
where
\begin{align}\label{c2}
[\hat d_\psi]_{ij}=c_1[\hat d]_{ij}+c_2{\text{Tr}}[\hat d]\,\delta_{ij},~~~~~~~[\hat d'_\psi]_{ij}=c'_1[\hat d']_{ij}+c'_2{\text{Tr}}[\hat d']\,\delta_{ij}
\end{align}
and
\begin{align}\label{dhat}
\hat d&=\frac12\left(\xi^\dagger d\xi^\dagger+\xi d^\dagger\xi\right)=d-\Pi d\Pi-\frac12\left(\Pi^2 d+ d\Pi^2\right)+\cdots\\\no
\hat d'&=\frac{i}2\left(\xi^\dagger d\xi^\dagger-\xi d^\dagger\xi\right)=\Pi d+d\Pi+\cdots.
\end{align}
Similarly to $b$, the RG evolution of $d$ down to $\Lambda\sim M_\chi$ and the result of the matching between \eqref{Lag} and the chiral quark model is encoded in $c_1,c_2$ and $c_1',c_2'$.

From \eqref{dipNRQM}, and setting the pion matrix to its vacuum expectation value $\langle\xi\rangle=1$, we readily see that the EDMs of the constituent quarks appearing in \eqref{NRQMtree} are related to the bare EDMs of \eqref{Lag} via
\ba
[d_\psi]_{ij}=c_1d_{ij}+c_2{\text{Tr}}[d]\delta_{ij}. 
\ea
The presence of two coefficients stems from the fact that the traceless and trace parts of $d$ transform as two independent representations of the unbroken $SU(3)_V$, and contribute differently to physical observables, as already emphasized in Section \ref{sec:largeN}. Following our prescription \eqref{NRQMtree}, and employing the definition \eqref{nEDM}, we have   
\ba\label{kappaNRQM}
g_{Tn}^u=-\frac13 c_1+c_2,~~~~~g_{Tn}^d=\frac43 c_1+c_2,~~~~~g_{Tn}^s=c_2.
\ea
In terms of fermions $\psi'_L=\xi\psi_L$, $\psi'_R=\xi^\dagger\psi_R$ transforming under $SU(3)_L\times SU(3)_R$ as $q_L$, $q_R$ in \eqref{Lag}, $c_2$ in \eqref{dipNRQM} is the coefficient of an operator with two insertions of $ U$, which is dual to the chiral condensate. That coefficient therefore arises from matching the chiral quark model to the class of operators \eqref{LagW} discussed in Section \ref{sec:RG}. Hence, as anticipated in Section \ref{sec:RG}, at leading order in the bare quark masses $g_{Tn}^s$ is {\emph{entirely}} induced by dangerous irrelevant operators like \eqref{LagW}. 
Of course we cannot reliably calculate $c_{1,2}$ because that would require a non-perturbative calculation, but we can estimate them invoking naturalness arguments as done in \cite{Manohar:1983md}: the size of $c_{1,2}$ should be comparable to, and certainly not much smaller than, the typical radiative corrections derived within the EFT. Let us look at the corrections to $c_2$ proportional to $c_1$. 

Corrections proportional to $c_1$ that involve only quark-gluon loops are analogous to those that one would find in ordinary QCD. The dominant effect of that type arises at 3 loops from corrections analogous to that shown in the first diagram of Fig. \ref{FigDI}. Yet, in the chiral quark model that kind of diagrams are calculable because the constituent mass $m_\psi$ acts as a natural IR regulator and avoids the problem we encountered in perturbative QCD. In fact, the 3-loop contribution is finite \cite{Grozin:2008nw,Ema:2022pmo}. Taking a common constituent mass $m_\psi$, using ref. \cite{Ema:2022pmo} we find
\ba\label{3-loop}
\left. \delta c_2\right|_{\text{gluons}}= c_1\frac{5(8\zeta(3)-7)}{72}\left(\frac{g^2}{4\pi^2}\right)^3=c_1\,0.18\left(\frac{g^2}{4\pi^2}\right)^3\sim c_1\,10^{-4}.
\ea
Note that this result is numerically consistent with a naive large $N_c$ estimate, which for $N_c=3$ reads $\delta c_2\sim\frac{1}{N_c}(g^2N_c/16\pi^2)^3=0.14(g^2/4\pi^2)^3$. In principle, \eqref{3-loop} could be sizable if the coupling was strong. However, the $g$ indicated by \cite{Manohar:1983md} is quite small, and so such correction turns out to be very tiny, as shown in the numerical estimate of \eqref{3-loop}.

\begin{figure}[t]
\begin{center}
\includegraphics[width=.35\textwidth]{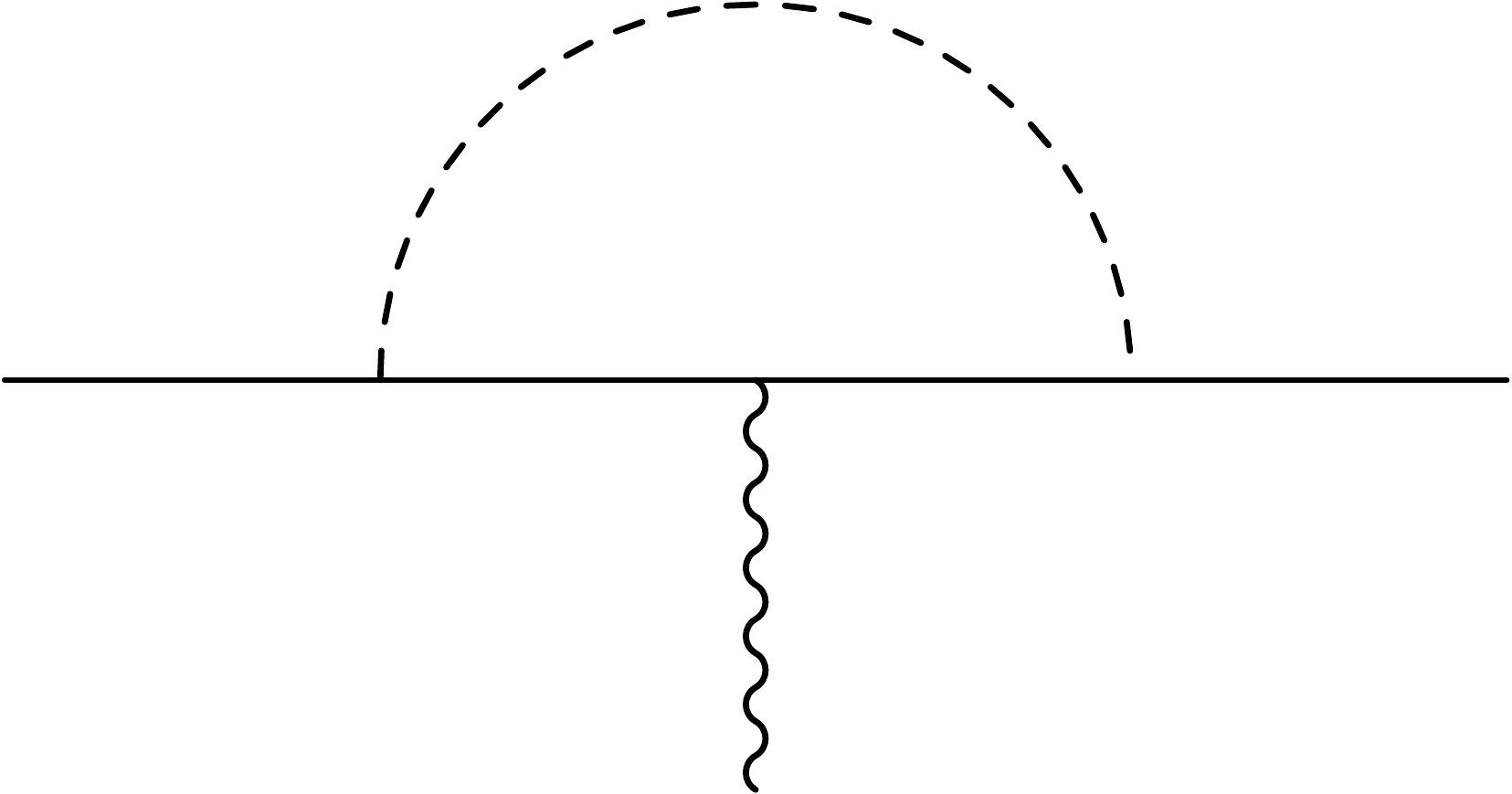}
\caption{\small 1-loop diagram describing the additive renormalization of $c_2$ in the chiral quark model (see \eqref{c2}).}\label{FigchiQM}
\end{center}
\end{figure}

There are also corrections to $c_2$ that involve virtual mesons, which are characteristic of the chiral quark model. At 1-loop there is a single diagram that can induce a non-vanishing $c_2$ from $c_1\neq0$. It is shown in Fig. \ref{FigchiQM} and has a virtual strange of mass $ m_{\psi_s}=m_\psi+{\cal O}(m_s)\approx540$ MeV\footnote{Here we included a formally subleading ${\cal O}(m)$ correction to the constituent mass because it is numerically important. Technically, it is parametrized by a $-\overline{\psi}\hat m\psi$ term in \eqref{chiQM}, where $\hat m$ is a CP-even combination of $\xi^\dagger m\xi^\dagger$ analogous to the one on the left in \eqref{dhat} with $m$ replacing $d$.} and a virtual Kaon of mass $M_K\approx494$ MeV. Its flavor structure 
\ba
(T^ad_\psi T^a)_{ij}=\frac12{\text{Tr}}[d_\psi]\delta_{ij}-\frac16 [d_\psi]_{ij}
\ea
introduces the desired dependence of $d_\psi$ on ${\text{Tr}}[d_\psi]$. The diagram is UV-divergent, and the divergence is cancelled by a local counterterm in \eqref{dipNRQM}. If we assume a sharp cutoff at the maximum scale $M_\chi$, taken for simplicity $\gg  m_{\psi_s},M_K$, the size of $c_2$ would be  
\ba\label{c2NDA}
\left.\delta c_2\right|_{\text{NDA}}\sim c_1g_A^2\frac{M_\chi^2}{16\pi^2f_\pi^2}\sim \frac{c_1}{N_c}.
\ea
Recalling \eqref{kappaNRQM} and \eqref{Mchi} we see that the resulting form of $g_{Tn}^s/g_{Tn}^{u,d}$ is parametrically consistent with the large $N_c$ scaling presented in Section \ref{sec:largeN}.

Yet, one might say Eq. \eqref{c2NDA} is an overestimate since it is plausible that the meson dynamics in the chiral quark model is in fact softer than in the more familar chiral Lagrangian. This is because if the effective meson coupling $p^2/f_\pi^2$ could grow up to the maximal value $\sim16\pi^2/\sqrt{N_c}$ as in the chiral Lagrangian, then radiative corrections would be significant and our predictions would depart significantly from those of a naive NRQM, which are known to be rather accurate. If \eqref{chiQM} combined with \eqref{wavefunN} have to reproduce the successful predictions of the non-relativistic quark model, then the pion loops cannot be too significant. After all, QCD should be better and better approximated by a pure quark-gluon Lagrangian as the momentum scale increases. To mimic this feature we may imagine the loop of Fig. \ref{FigchiQM} gets saturated at $<M_\chi$. The softest scale that allows \eqref{chiQM} to make sense is $ m_{\psi_s}$. Therefore, in order to obtain a more conservative estimate of the size of the radiative corrections to $c_2$ we perform a leading log approximation of Fig. \ref{FigchiQM} using Dimensional Regularization, obtaining
\ba\label{loopGM}
\left.\delta c_2\right|_{\text{LL}}
=c_1\frac{g_A^2}{16\pi^2f_\pi^2}\left[\frac{7}{12} m_{\psi_s}^2\ln\frac{\mu^2}{ m_{\psi_s}^2}+\frac12M_K^2\ln\frac{\mu^2}{M_K^2}\right].
\ea
Taking $g_A\sim0.75$, $ m_{\psi_s}\sim 540$ MeV and $\mu=1$ GeV, the result is numerically $\left.\delta c_2\right|_{\text{LL}}\sim0.2\,c_1$, and thus not far from our previous guess \eqref{c2NDA}. The parametric $N_c$ dependence is the same since $g_A={\cal O}(N_c^0)=m_{\psi_s}$. We explicitly presented also the leading non-analytic correction in $M_K^2\propto m_s$ because it is numerically comparable to the first term in \eqref{loopGM}.

Since the constituent quark mass is related to the condensate as $\langle\overline q_iq_j\rangle\sim\delta_{ij}m_\psi f_\pi^2$, the parametric dependence displayed in the first term in \eqref{loopGM} is precisely the same expected from a matching to operators like \eqref{LagW}. (The only ingredient that was missing in Section \ref{sec:RG} is the $1/N_c$ suppression.) Furthermore, the non-analytic contribution $M_K^2\ln M_K^2$ shown in the second term in \eqref{loopGM} comes along with (incalculable) counterterms of order $M_K^2/(4\pi f_\pi)^2$. The latter are evidently the result of a matching between \eqref{chiQM} and \eqref{LagW2}, and their coefficients should not be much different from the non-analytic terms, and hence compatible with NDA. Overall, we may say that the naturalness argument employed above to estimate \eqref{c2NDA}, or the more conservative leading log calculation in \eqref{loopGM}, represent indirect evidence that the NDA analysis of Section \ref{sec:RG} is accurate.

In conclusion, a closer look at the NRQM confirms the expectations from Section \ref{sec:RG} and the qualitative reliability of NDA arguments. More quantitatively, the present model suggests
\ba\label{ratioQM}
\frac{|g_{Tn}^s|}{|g_{Tn}^{u,d}|}\sim0.1\div0.3
\ea
as indicated by \eqref{c2NDA} and \eqref{loopGM}. The strange tensor charge cannot be neglected even in a prescription like \eqref{NRQMtree} in which the impact of the constituent strange identically vanishes at the hadronic scale.

\section{Chiral perturbation theory}
\label{sec:chiralPT}

In this section we will use chiral perturbation theory to validate the NDA estimates of Section \ref{sec:RG} and the approach of Section \ref{sec:NRQM}.

In a heavy baryon EFT approach \cite{Jenkins:1990jv} the neutron EDM can be systematically expanded as a function of the electromagnetic coupling and
\ba\label{HBEFTpar}
\frac{M_\Pi}{\Lambda_\chi},~~~\frac{\delta M}{\Lambda_\chi}
\ea
where $M_\Pi$ are the masses of the meson octet, $\delta M$ the mass splitting between an excited baryon and the neutron in the exact $SU(3)_V$ limit, and\footnote{In this section we work with $N_c=3$ and do not keep track of the large $N_c$ scalings.}
\ba
\Lambda_\chi\equiv 4\pi f_\pi.
\ea

The corrections controlled by $M_\Pi$ and electromagnetism measure the departures from the exact flavor $SU(3)_{V}$ limit. The dominant effects are those proportional to the strange quark mass, and in the following we will ignore all the others. In that limit the pion triplet remains massless whereas $M_\eta^2=\frac43M_K^2\propto m_s\Lambda_\chi$. Quantitatively, the breaking of $SU(3)_{V}$ is quite sizable, say of order $M_K^2/\Lambda_\chi^2\sim20\%$, and for this reason a perturbative expansion in $M_K/\Lambda_\chi$ may not always be trustable. We will nevertheless restrict our analysis to the leading non-trivial order in $M_K/\Lambda_\chi$, and comment on the impact of higher order corrections at the end.

Baryons with $\delta M\gtrsim\Lambda_\chi$ as well as mesons with masses larger or of order $\Lambda_\chi$ are integrated-out, and their effect is captured by local counterterms. The baryons to be included in the EFT as dynamical states are the baryon octet, the baryon decuplet, as well as several other excitations with $\delta M$ smaller than $\Lambda_\chi$. The baryon octet of course is defined by $\delta M=0$, the decuplet has $\delta M\sim200$ MeV, and so on. The splittings among the components of a $SU(3)_V$ multiplet, due to $SU(3)_V$ breaking, are of order $m_s$. They give rise to corrections to $d_n$ proportional to $m_s/M_\Pi\sim (m_s/\Lambda_\chi)^{1/2}$, which are higher order in the chiral expansion and will be neglected. Therefore, in our leading order analysis the baryons form degenerate $SU(3)_V$ multiplets.

\subsection{Dipoles at leading order}

In the heavy baryon EFT the baryon octet has approximately constant 4-velocity $v^\mu$, slightly perturbed by interactions with other hadrons. It is described by a traceless matrix-valued field 
\ba\label{PsiB}
\Psi=\left(
\begin{matrix} 
\frac{\Sigma_0}{\sqrt2}+\frac{\Lambda}{\sqrt6} & \Sigma_+ & p\\\Sigma_- & -\frac{\Sigma_0}{\sqrt2}+\frac{\Lambda}{\sqrt6} & n \\ \Xi_- & \Xi_0 & -\frac{2\Lambda}{\sqrt6}\end{matrix}\right)
\ea
transforming under $SU(3)_{V}$ as $\Psi\to h\Psi h^\dagger$ and satisfying $v_\mu\gamma^\mu  \Psi=\Psi$ \cite{Jenkins:1990jv}. The dominant interactions with the light mesons are given by:
\begin{align}\label{LO}
{\cal L}_{0}
&=\frac{f_\pi^2}{4}{\text{Tr}}\left[\partial_\mu U\partial^\mu U^\dagger\right]-B{\text{Tr}}\left[m U^\dagger+m^\dagger U\right]\\\no
&+{\text{Tr}}\left[\overline{\Psi}v^\mu i\nabla_\mu\Psi\right]+(D+F){\text{Tr}}\left[\overline\Psi \gamma^\mu\gamma_5 A_\mu\Psi\right]+(D-F){\text{Tr}}\left[\overline\Psi \gamma^\mu\gamma_5 \Psi A_\mu\right],
\end{align}
where the meson octet field has been introduced in Section \ref{sec:NRQM}. In the second line of \eqref{LO}, $\nabla_\mu\Psi=\partial_\mu\Psi+[V_\mu,\Psi]$ is the appropriate $SU(3)_L\times SU(3)_R$-covariant derivative, and $V_\mu,A_\mu$ are the same as shown in \eqref{VA}. The explicit value of the parameter $B$ will not be relevant for us. Its main effect is introducing masses for the light mesons. The quantities $D,F$ are obtained by matching the predictions of the EFT with observed semi-leptonic baryon decays. At tree-level one has $D\sim0.8$, $F\sim0.5$ (at 1-loop $D\sim 0.56$, $F\sim 0.37$). 

In addition to \eqref{LO} there are interactions involving higher powers of derivatives or insertions of $m$, couplings to heavy baryons with $\delta M>0$, as well as operators proportional to the CP-violating coupling $d$. At leading order in the chiral expansion, there are two classes of baryon-octet interactions containing a single insertion of $d$:
\begin{align}\label{dip}
{\cal L}_{\text{dip}}
=&-\frac{i}{2}\left\{\kappa_1{\rm Tr}[\hat d\overline{\Psi} \sigma^{\mu\nu}\gamma_5\Psi]+\kappa_2{\rm Tr}[\overline{\Psi} \sigma^{\mu\nu}\gamma_5\hat d\Psi]+\kappa_3{\text{Tr}}[\hat d]{\rm Tr}[\overline{\Psi} \sigma^{\mu\nu}\gamma_5\Psi]\right\}F_{\mu\nu}\\\no
&-\frac{1}{2}\left\{\kappa_1'{\rm Tr}[\hat d'\overline{\Psi} \sigma^{\mu\nu}\Psi]+\kappa_2'{\rm Tr}[\overline{\Psi} \sigma^{\mu\nu}\hat d'\Psi]+\kappa_3'{\text{Tr}}[\hat d']{\rm Tr}[\overline{\Psi} \sigma^{\mu\nu}\Psi]\right\}F_{\mu\nu},
\end{align}
with $\hat d$ and $\hat d'$ the same quantities defined in \eqref{dhat}. The couplings $\kappa_1,\kappa_2,\kappa_3$ parametrize the baryonic EDMs at zeroth order in the chiral expansion. We collect the leading order expression of the EDMs in Appendix \ref{app:chiralPT}. In particular, the EDM of the neutron reads
\ba
d^{\text{LO}}_{n}=\kappa_1 d_s+\kappa_2 d_d+\kappa_3(d_u+d_d+d_s).
\ea
To make contact with our earlier notation, we observe that the leading order prediction for the tensor charges in \eqref{nEDM} are
\begin{align}\label{LOcoeff}
g_{Tn,\text{LO}}^u&=\kappa_3\\\no
g_{Tn,\text{LO}}^d&=\kappa_2+\kappa_3\\\no
g_{Tn,\text{LO}}^s&=\kappa_1+\kappa_3.
\end{align}
The parameters $\kappa_3$, $\kappa_2+\kappa_3$, or $\kappa_1+\kappa_3$ are all expected to be of order unity according to NDA. Yet, the particular combination 
\ba\label{holds}
g_{Tn,\text{LO}}^u+g_{Tn,\text{LO}}^d+g_{Tn,\text{LO}}^s=\kappa_1+\kappa_2+3\kappa_3
\ea
is somewhat special since, in the absence of $SU(3)_V$ breaking, the condition $g_{Tn}^u+g_{Tn}^d+g_{Tn}^s=0$ is RG-invariant. The reason is that when such relation holds the trace ${\text{Tr}}[d]$ decouples from \eqref{dip}, and only the traceless part survives. The RG-stability of the condition ${\text{Tr}}[d]=0$ is of course not a surprise. In fact, in the large $N_c$ expansion, terms proportional to ${\text{Tr}}[d]$ arise from closed fermion loops and must be suppressed, so in the exact $N_c=\infty$ limit the condition ${\text{Tr}}[d]=0$ must be stable, which is what we have just observed. What is remarkable is that in this section we are able to identify the special condition $\kappa_1+\kappa_2+3\kappa_3=0$ (previously shown in \eqref{SU3relation}) in an EFT without quarks, just based on basic considerations.

In the following we will be agnostic about the relative size of $\kappa_1,\kappa_2,\kappa_3$ and calculate the chiral corrections to \eqref{LOcoeff}.

\subsection{NLO corrections}

\begin{figure}[t]
\begin{center}
\includegraphics[width=.2\textwidth]{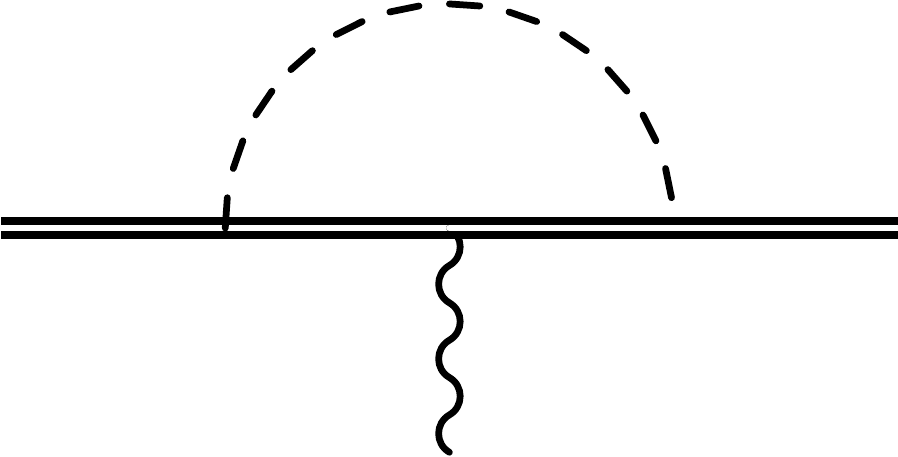}~~~~~~
\includegraphics[width=.2\textwidth]{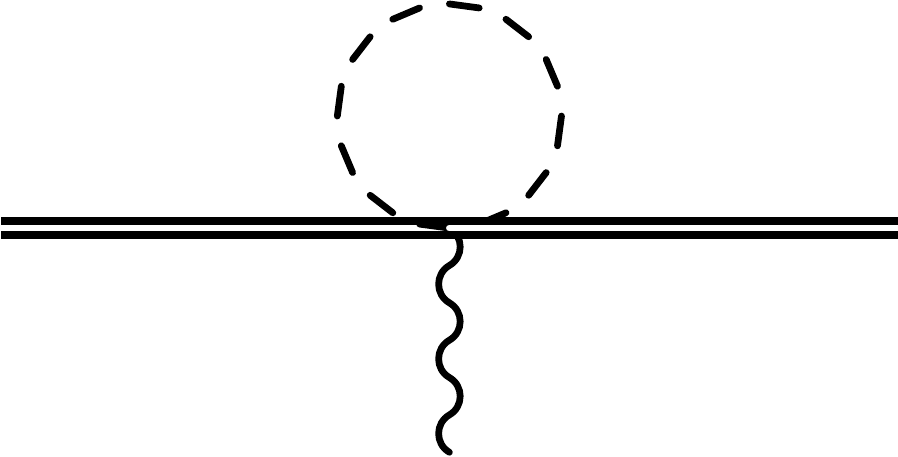}~~~~~~
\includegraphics[width=.2\textwidth]{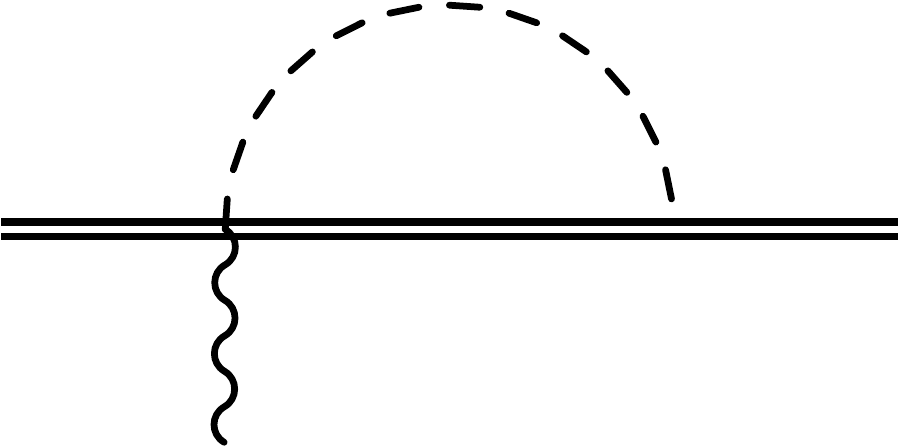}~~~~~~
\includegraphics[width=.2\textwidth]{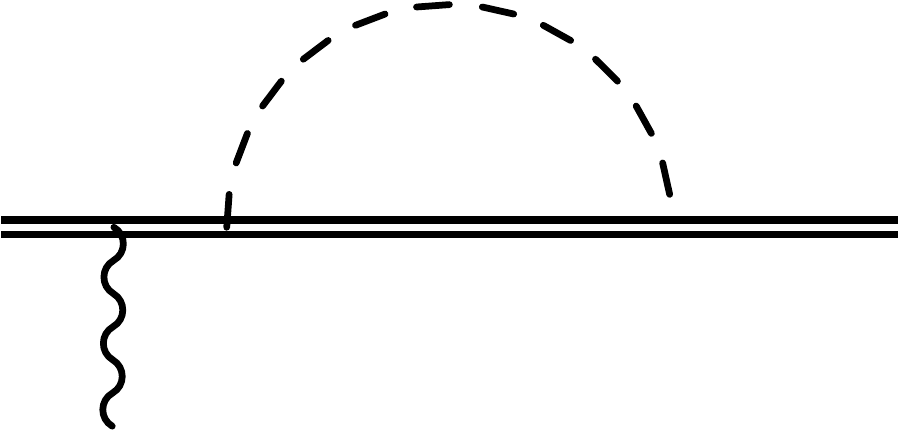}
\caption{\small Next to leading order contributions to the baryon dipole moments. Double lines indicate baryons, dashed lines refer to the light mesons and the wavy line to the external electric field.}\label{Fig}
\end{center}
\end{figure}

The next to leading order corrections in the chiral expansion are induced by the baryon-octet 1-loop diagrams of Fig. \ref{Fig}, involving one insertion of ${\cal L}_{\text{dip}}$ and the baryon-meson interactions from ${\cal L}_{0}$, plus corrections due to virtual heavier baryons. A simple power-counting analysis reveals that at NLO the baryonic dipoles acquire the following form
\ba\label{NLOgen}
d^{\text{NLO}}_{a'b'}=d^{\text{LO}}_{a'b'}+A^\Pi_{a'b'}\frac{M_\Pi^2}{16\pi^2f_\pi^2}\ln\frac{\mu^2}{M_\Pi^2}+B^\Pi_{a'b'}\frac{M_\Pi^2}{16\pi^2f_\pi^2},
\ea
with $a',b'$ indices labelling the baryons and in our approximation $\Pi=K,\eta$. The quantities $A^\Pi_{a'b'}$ are calculated from the diagrams of Fig. \ref{Fig} with virtual baryon octets and meson octets, and are fully determined in terms of the parameters $d$, $\kappa_1,\kappa_2,\kappa_3$ and $D,F$. On the other hand, loops involving heavier baryons introduce calculable functions of $\delta M/M_K$ and $\delta M/\mu$, which however depend on the unknown dipoles of the heavy baryons and light-heavy transition dipoles. The $M_K$-independent part of these contributions can be re-absorbed in the $SU(3)_{V}$ invariant parameters $\kappa_{1,2,3}$. The reminder is proportional to $M_K$ and is included, along with the unknown counterterms of order ${\cal O}(dm)$, into $B^\Pi_{a'b'}$. The latter are of course incalculable within our EFT, but NDA suggests $B^\Pi_{a'b'}$ should be numerically of order unity. 

Since $\ln \mu^2/M_K^2$ and $\ln \mu^2/M_\eta^2$ for the natural choice $\mu\sim M_n$ are of order unity, there are no large-log enhancements in \eqref{NLOgen}; hence, the fully calculable non-analytic corrections proportional to $A^\Pi_{a'b'}$ contribute comparably to the incalculable counterterms $B^\Pi_{a'b'}$. For this reason, in the absence of additional phenomenological input one cannot make a numerically accurate prediction of $d_n$. Nevertheless, qualitative conclusions can still be drawn, as we will see.

We calculated the diagrams of Fig. \ref{Fig} employing Dimensional Regularization. Details of the computation are presented in Appendix \ref{app:chiralPT}. Here we quote only the NLO expression for $g_{Tn}^s$ evaluated at $\mu=M_n$ (see \eqref{here3}):
\begin{align}\label{dotsandc}
g_{Tn,\text{NLO}}^s
&=\frac{M^2_K}{16\pi^2f_\pi^2}\ln\frac{M_n^2}{M_K^2}\left(\frac12-\frac{13}{18}D^2+\frac{15}{9} DF-\frac12 F^2\right)g_{Tn,\text{LO}}^u\\\no
&+\frac{M^2_K}{16\pi^2f_\pi^2}\ln\frac{M_n^2}{M_K^2}\left(\frac12-\frac{1}{18}(D+3 F)^2\right)g_{Tn,\text{LO}}^d\\\no
&+\left\{1+\frac{M_K^2}{16\pi^2f_\pi^2}\ln\frac{M_n^2}{M_K^2}\left(1+\frac{22}{9}D^2-\frac{10}{3}DF+4F^2\right)\right.\\\no
&\left.~~~+\frac{M_\eta^2}{16\pi^2f_\pi^2}\ln\frac{M_n^2}{M_\eta^2}\left(\frac23+\frac{1}{6}(D-3F)^2\right)\right\}g_{Tn,\text{LO}}^s\\\no
&+B_{nn,s}^K\frac{M_K^2}{16\pi^2f_\pi^2}.
\end{align}
Importantly, we see that $g_{Tn}^s$ mixes at order $M_K^2/\Lambda_\chi^2=0.2$ with $g_{Tn}^{u,d}$ because of loops with a virtual $\Sigma$ baryon and a Kaon.\footnote{The relevance of the $\Sigma,K$ loops in the study of the strange chromo-electric dipole was emphasized by \cite{Khatsimovsky:1987bb,Hisano:2004tf}. One finds analogous sizable corrections to the matrix element of axial currents \eqref{matrixelementA}, see for instance \cite{Savage:1996zd}.} Quantitatively, evaluating \eqref{dotsandc} numerically with $M_K=494$ MeV, $f_\pi=92$ MeV and the values of $D,F$ extracted at tree-level, we find
\begin{align}\label{dotsandc1}
g_{Tn,\text{NLO}}^s
&=0.14\, g_{Tn,\text{LO}}^u+0.048\,g_{Tn,\text{LO}}^d+1.7\, g_{Tn,\text{LO}}^s+0.18\,B_{nn,s}^K.
\end{align}
(Using the 1-loop values of $D,F$ only the second entry would be appreciably affected, going from $0.048$ to $0.081$.) So, even if for some accident the leading order $|g_{Tn,\text{LO}}^s|$ term is negligible, the strange tensor charge receives corrections proportional to the valence quark tensor charges of the natural size $M_K^2/\Lambda_\chi^2$. Barring unnatural cancellations we thus conclude that
\ba\label{chiPTgs}
\frac{|g_{Tn}^s|}{|g_{Tn}^{u,d}|}\gtrsim0.1. 
\ea
We interpreted this as a lower bound because in reality we saw in Sections \ref{sec:largeN} and \ref{sec:NRQM} that $g_{Tn,\text{LO}}^s$ is typically non-vanishing. Note also that the numerical size of the non-analytic $m_s\ln m_s$ term in Eq. \eqref{loopGM} is consistent with the one presented in \eqref{dotsandc}. This proves that the naturalness considerations of Section \ref{sec:NRQM} are sensible. 

Let us finally comment on the impact of higher order corrections on \eqref{NLOgen}. We have already mentioned that in the real world a chiral expansion may not be fully trustable because $M_K^2/\Lambda_\chi^2=0.2$ is not very small. What this means in practice is that our approach may suffer from large numerical uncertainties due to higher order corrections. What can our analysis teach us, then? Well, for sure $g_{Tn}^s$ is a complicated function of $M_K/\Lambda_\chi$, of which we have found just the first few terms in a hypothetical $M_K/\Lambda_\chi\ll1$ limit. And the very fact that in such a limit the strange tensor charge does depend on $m_s$ 
allows us to draw an important robust conclusion, which holds irrespectively of whether $M_K/\Lambda_\chi$ can or cannot be considered a small parameter: a ratio $|g_{Tn}^s/g_{Tn}^{u,d}|$ much smaller than shown in \eqref{chiPTgs} can {\emph{only}} be the result of a $m_s$-dependent cancellation among a priori unrelated quantities, namely the leading order $m_s$-independent value $g_{Tn,\text{LO}}^s$ and the chiral corrections. The presence of such a cancellation cannot be excluded by our analysis, but it is fair to say that it would be very surprising. 

\section{Discussion}
\label{sec:conclusions}

An accurate determination of how the dipole moments of the light quarks affect the neutron EDM $d_n$ is essential to the correct interpretation of the experimental measurements of $d_n$ within a beyond the Standard Model hypothesis.\footnote{In the Standard Model, the dominant contribution to $d_n$ arises from a double insertion of the Fermi interactions dressed by long-range effects. According to NDA we expect it to be of order $d_n/e\sim J({f_\pi}/{v})^4/M_n\sim10^{-32}{\text{cm}}$, where $J\sim3\times10^{-5}$ is the Jarlskog invariant and $v\sim246$ the electroweak scale. The contribution arising from quark dipoles is instead chirally suppressed and numerically smaller by roughly a factor $10^{-2}$ \cite{Khriplovich:1985jr,Czarnecki:1997bu}. Hence, our results do not impact the Standard Model prediction of $d_n$ substantially. We thank T. Cohen for asking this important question.} In this paper we approached this interesting question, focusing in particular on the strange quark contribution, see \eqref{nEDM}.

We approached the problem using perturbative QCD and the large $N_c$ expansion (combined with NDA arguments), a concrete modelling of the constituent quark dynamics (combined with naturalness considerations), and finally chiral perturbation theory. While none of these tools can on its own determine the tensor charges precisely, the strength of our assessment resides in the fact that all our predictions turn out to be perfectly consistent with each other, both qualitatively as well as quantitatively.

Our main results can be summarized as follows. The strange tensor charge $g_{Tn}^s$ is controlled by IR-sensitive diagrams of order $1/N_c$ that contain either insertions of the quark masses or insertions of the quark condensate. At around the GeV such diagrams are unsuppressed and induce a $g_{Tn}^s$ of order $0.1\div0.3$ of the tensor charges $g_{Tn}^{u,d}$ of the valence quarks. Despite this mild suppression, and the fact that the sign of $g_{Tn}^s$ remains unknown, our results demonstrate that $g_{Tn}^s$ is absolutely critical to scenarios beyond the Standard Model, a possibility that is overlooked in most phenomenological studies. Indeed, in many extensions --- specifically those in which the quark EDMs are proportional to the mass of the corresponding quark --- the dominant contribution to $d_n$ is due to the strange quark dipole as soon as $|g_{Tn}^s/g_{Tn}^{u,d}|$ is larger than about $0.02\div0.05$, which is what our results actually indicate.

At first, the existence of an unsuppressed $g_{Tn}^s$ might seem to contradict the non-relativistic quark model, where the strange is naively expected to be irrelevant, being a sea quark. However, this is not the case. Because in such a picture the dynamical degrees of freedom are {\emph{constituent}} quarks, before drawing any robust conclusion about $g_{Tn}^i$ one should first establish a mapping between the bare quark dipoles and the constituent quark dipoles. The chiral quark model of Georgi and Manohar represents a simple enough framework in which such a matching can be concretely approached. Consistently with our diagrammatic arguments, in that model $g^s_{Tn}$ arises from $1/N_c$-suppressed contributions proportional to the bare quark masses and the vacuum condensate, and its numerical size is naturally of the order mentioned above.

Perhaps the most convincing piece of evidence in support of our claims comes from our calculation in heavy baryon EFT. The part of $g_{Tn}^s$ that depends non-analytically on the quark masses can be unambiguously extracted from chiral perturbation theory, at least as long as $m_s/(4\pi f_\pi)$ can be considered a small parameter. A next-to-leading order analysis reveals that those contributions are of the natural size $M_K^2/(4\pi f_\pi)^2$, in agreement with the chiral quark model as well as with NDA and large $N_c$ expectations. Barring accidental cancellations among a priori unrelated quantities, this implies that $|g_{Tn}^s/g_{Tn}^{u,d}|$ cannot be smaller than $0.1$. While our analysis rests on the assumption that $m_s/(4\pi f_\pi)$ be a small parameter, the conclusion that $g_{Tn}^s$ does depend on $m_s$, and that among the $m_s$-dependent contributions there are some of order $0.1$, are of course robust. As a result we can firmly establish that a hypothetical strange tensor charge much smaller than $0.1$ can only be the result of a $m_s$-dependent cancellation, irrespective of whether the chiral expansion in $m_s/(4\pi f_\pi)$ is accurate or not.

The pattern that emerges is qualitatively the same observed for the form factors of the axial vector operator \eqref{matrixelementA}. Even in that case $|g_{An}^s/g_{An}^{u,d}|$ is in the $10\%$ ballpark, as indicated by a multitude of independent data. In fact, virtually all of the results we derived here apply to $g_{An}^i$ as well.\footnote{With the exception of the comment around \eqref{matrixelementA}, which however does not appear relevant at the scales of interest.} In other words, we found no reason for $|g_{Tn}^s/g_{Tn}^{u,d}|$ to be ``special".

Our results are in conflict with the numerical results of the lattice QCD investigations available at present, which find a strikingly small $g_{Tn}^s$, see \eqref{latticeQCD}. The $1/N_c$ suppression cannot justify the discrepancy, as if large $N_c$ was the true answer, then one would also anticipate $g_{Tn}^d/g_{Tn}^u\approx-1$, which instead appears to be badly violated by the measurement quoted in the first relation in \eqref{latticeQCD}. At present the disagreement between \eqref{latticeQCD} and our results still puzzles us, but we feel we should briefly comment about it. 

First, we have no reason to doubt the validity of the lattice simulations used in \cite{Bhattacharya:2016zcn,Gupta:2018lvp,Park:2025rxi}, and we will therefore present our considerations under the assumption that the discretized strange tensor charge is actually negligible. Because \eqref{latticeQCD} is suggestively close to the prediction of a naive NRQM, maybe this is one of those instances in which the NRQM turns out to describe QCD much better than ``it should". 
Second, and regardless of that, it is well known that even in the presence of precise measurements of the lattice matrix elements, one is still a long way to an evaluation of the parameters $g_{Tn}^i$ that appear in the continuum Lagrangian \eqref{Lag}. The calculations of \cite{Bhattacharya:2016zcn,Gupta:2018lvp,Park:2025rxi}, for example, are performed at finite lattice spacing $a$, subsequently matched to a continuum RI-MOM scheme at a scale of order $\Lambda$, then converted into Dimensional Regularization at that scale, and finally ran up to $\Lambda_{\text{UV}}=2$ GeV. We would expect this matching to be impacted by the finite threshold effects discussed in Section \ref{sec:RG}. However, we do not see any trace of the associated diagrams in the references used in \cite{Bhattacharya:2016zcn,Gupta:2018lvp,Park:2025rxi}. More generally, the extraction of a $|g_{Tn}^s/g_{Tn}^{u,d}|$ smaller than $1\%$, as in \eqref{latticeQCD}, requires understanding the extrapolation to the continuum with a precision that is at least of the same order and, in our opinion, such an effort cannot escape the evaluation of the diagrams of Section \ref{sec:RG} at a scale $\Lambda$ numerically close to the GeV, since the matching ideally must be performed within a window $M_h\ll \Lambda \ll1/a$ \cite{Martinelli:1994ty}. Given that the QCD coupling is inevitably largish at that scale, even 3-loop diagrams mixing the strange and valence quark dipoles turn out to be crucial, especially for $g_{Tn}^s$, which we have argued is dominated by such contributions. Finally, our NLO analysis in heavy baryon EFT firmly establishes that a hypothetical suppression of $|g_{Tn}^s/g_{Tn}^{u,d}|$ can only be the result of a $m_s$-dependent cancellation. Hence, a careful analysis of the $m_s$ dependence of $g_{Tn}^s$ represents a precious consistency check for future numerical investigations.

\section*{Acknowledgments}

We thank T. Bhattacharya and R. Gupta for discussions about their simulation, M. Luty for sharing his wisdom on large $N_c$ QCD, and M. Pospelov for sharing his view on this problem. This work was partly supported by the Italian MIUR under contract 202289JEW4 (Flavors: dark and intense), the Iniziativa Specifica “Physics at the Energy, Intensity, and Astroparticle Frontiers” (APINE) of Istituto Nazionale di Fisica Nucleare (INFN), and the European Union’s Horizon 2020 research and innovation programme under the Marie Sklodowska-Curie grant agreement No 860881-HIDDeN.

\appendix

\section{More on large $N_c$}

 \subsection{One-body matrix elements}
 \label{app:onebody}
 
In this appendix we collect results useful for the analysis of Section \ref{sec:largeN}, and are a straightforward application of \cite{Luty:1994ua}.

We consider the matrix element of 1-body operators of the form 
$$
\left\{W J_k\right\}=W^i_{j}\alpha_{i\alpha}^\dagger[\sigma_k]^\alpha_\beta\alpha^{j\beta},~~~~~~~\left\{W\right\}=W^i_{j}\alpha_{i\alpha}^\dagger\alpha^{j\alpha},
$$
with $W$ hermitian in flavor space. We can write
\ba
W=\frac13{\text{Tr}}[W]{\bf 1}+W_a\lambda_a.
\ea
In the text we presented some result for $W=d$.

Our normalized neutron of spin $+1/2$ is described by \eqref{normn}, where $\alpha_{u\pm/d\pm}^\dagger$ are the creation operators for up/down quarks of spin $\pm1/2$, and $|0)$ is an appropriate vacuum state. The normalization immediately follows from the standard definitions $[\alpha^{i\alpha},\alpha^{j\beta}]=0$ and $[\alpha^{i\alpha},\alpha_{j\beta}^\dagger]=\delta^\alpha_\beta\delta^i_j$. We find
\begin{align}
(n|n)&=C_n^2\frac18(N_c+3)(N_c+1)\,\Gamma^2\left(\frac{N_c+1}{2}\right)=1.
\end{align}
It is straightforward to see see that
\begin{align}\label{formulaGen}
(n|\left\{J_3\right\}|n)&=1\\\no
(n|\left\{\lambda_3J_3\right\}|n)&=-\frac{N_c+2}{3}\\\no
(n|\left\{\lambda_8J_3\right\}|n)&=\frac{1}{\sqrt{3}},
\end{align}
and
\begin{align}\label{formulaGen}
(n|\left\{{\bf 1}\right\}|n)&=N_c\\\no
(n|\left\{\lambda_3\right\}|n)&=-1\\\no
(n|\left\{\lambda_8\right\}|n)&=\frac{1}{\sqrt{3}}N_c.
\end{align}
With these simple formulas one can recover the expressions presented in Section \ref{sec:largeN}.

\subsection{The soliton model}
\label{app:soliton}

At large $N_c$ baryons can be viewed as topologically stable solitons of a theory of mesons \cite{Witten:1979kh,Witten:1983tx}, thus justifying the qualitative success of the approach pioneered by Skyrme \cite{Skyrme:1961vq}. We will now show that, at leading order in $1/N_c$, the soliton model makes a very sharp prediction regarding the relative size of the $g_{Tn}^i$'s.\footnote{The soliton model was previously employed in refs. \cite{Dixon:1990cq,Salomonson:1991ar} to estimate the $\bar\theta$-dependence of $d_n$. However, as far as we know, no analysis of the quark EDMs has ever been performed before (see however \cite{Olness:1992zb}).}

Imagine we are able to derive an EFT for the light meson octet by integrating out all hadrons of masses $M_\chi\gtrsim 4\pi f_\pi/\sqrt{N_c}$. The result is a chiral Lagrangian for the pion field $U$ with an infinite number of derivative interactions and a linearly realized $SU(3)_V$ flavor symmetry. In that language baryons are described by space-dependent configurations of the pion matrix, of the form
\ba\label{Sigma0}
U_0({\bf r})=e^{i F(|{\bf r}|)\lambda_k\hat {\bf r}_k},
\ea
where $\lambda_k$ are the Gell-Mann matrices associated to isospin $SU(2)_V\subset SU(3)_V$ (with $k=1,2,3$) and $F(|{\bf r}|)$ is a function found by extremizing the equations of motion and minimizing the energy. The typical size of the soliton is set by the dynamical scale, $\sim1/M_\chi={\cal O}(N_c^0)$, whereas its energy is of order $4\pi f_\pi^2/M_\chi={\cal O}(N_c)$.

The background solution \eqref{Sigma0} breaks translations, rotations and $SU(3)_V$ transformations. Yet, hypercharge and the combined action of ordinary rotations and isospin are conserved. We are interested in properties of baryons at rest, so we can ignore the motion of the center of mass. The seven collective coordinates associated to the remaining symmetry breaking pattern, $SU(2)_J\times SU(3)_V\to SU(2)_{J+V}\times U(1)_Y$, parametrize the coset $SU(3)/U(1)$ and are described by a time-dependent unitary matrix $A=A(t)$. The relevant fluctuations around the classical solution \eqref{Sigma0} are studied defining
\ba\label{Sigma}
U=AU_0 A^\dagger.
\ea
$SU(3)_V$ transformations act as $A\to VA$, with $V\in SU(3)_V$. Because $U_0(R{\bf r})= V^\dagger_S(R)U_0({\bf r}) V_S(R)$, with $ V_S(R)=e^{i \lambda_k\alpha_k}\in SU(2)_J$, ordinary rotations ${\bf r}\to R\,{\bf r}$ can equivalently be realized as $U_0({\bf r})\to U_0({\bf r})$ provided $A\to A V^\dagger_S(R)$. Therefore, the combined action of $SU(2)_J\times SU(3)_V$ is 
\ba\label{VR}
A\to VA\, V^\dagger_S(R).
\ea

To quantize the collective coordinates one proceeds as explained in \cite{Guadagnini:1983uv}\cite{Adkins:1984cf}\cite{Chemtob:1985ar}, which generalized to three flavors the approach of \cite{Witten:1983tx}\cite{Adkins:1983ya}. The idea is to consider the entire $SU(3)_V$ manifold and subject the matrix $A$ to an appropriate constraint that takes into account the redundancy $A\to Ae^{i\theta\lambda_8}$, valid for any time-dependent $\theta$. All the machinery necessary to our analysis has already been developed by those authors and need not be repeated here. We merely complete the picture by writing down the transformation properties of $A$ under the action of $P$, $CP$, and $T$. Analogously to what done for ordinary rotations (see around \eqref{VR}), we can define the transformation in such a way that $U_0({\bf r})$ in \eqref{Sigma} remains unchanged. The result is
\begin{align}
A(t)\xrightarrow[]{P}A(t),~~~~~~A(t)\xrightarrow[]{CP}A^*(t)\,(i\lambda_2),~~~~~~A(t)\xrightarrow[]{T}A^*(-t)\,(i\lambda_2).
\end{align}
For example, under $CP$ the familiar transformation of the NGB matrix is $U(x)\to U^*({\cal P}x)$. Because $ U_0^*(-{\bf r})= U_0^t({\bf r})=(i\lambda_2) U_0({\bf r})(i\lambda_2)^\dagger$, CP can be conveniently taken to act non-trivially only on $A$. Similar considerations can be made for $P$ and $T$.

The effective Lagrangian for $A$ in the absence of $d,\theta$ must be invariant under $P$, $CP$, and $T$ separately. It can be organized as an expansion in time derivatives of $A$, which must be small because our baryons are very heavy and move very little. On dimensional grounds we expect the time derivatives to appear in the combination $\partial_0/M_B$, where $M_B$ is the baryon mass. Because $M_B\propto N_c$, we see that our expansion in $1/N_c$ translates in this language into an expansion in time derivatives of $A$. The effect of the couplings $d,\theta$ can be easily taken into account applying the usual rules of effective field theories, treating them as spurions transforming appropriately under flavor $SU(3)_V$ as well as $P,CP,T$. Interestingly, neglecting derivatives of $A$, and compatibly with $SU(3)_V\times SU(2)_J$, there is a unique operator that can describe a spin $1/2$ baryon EDM:
\ba\label{EDMB}
{\bf d}_B=c\,{\text{Tr}}[A^\dagger d A\lambda_3],
\ea
with $c={\cal O}(N_c)$ a numerical factor. The presence of a unique operator is to be compared with considerations based solely on $SU(3)_V$, that allow 3 independent parameters (see Section \ref{sec:chiralPT}).

The expression \eqref{EDMB} is completely general and does not rely on any specific form of the meson EFT. Yet, just to be concrete we mention that the mesonic operator with the lowest number of derivatives and the correct quantum numbers to contribute to the $P$, $CP$, $T$-violating hamiltonian ${\cal H}_d=-{\cal L}_d$ is
\begin{align}\label{opd}
-\Delta {\cal L}_d
&= c'\left({\text{Tr}}[d  U^\dagger D_\mu U D_\nu  U^\dagger]-{\text{Tr}}[d^\dagger U D_\nu  U^\dagger D_\mu U]\right)i\widetilde F^{\mu\nu},
\end{align}
with $c'\sim f_\pi^2/M_\chi$ by naive dimensional analysis. Evaluating $H_d=\int d{\bf r}\,{\cal H}_d$ with $ U$ as in \eqref{Sigma} and an electric field $E_{\text{el}}$ along the 3-direction, one finds $\Delta H_d=-\Delta c\,{\text{Tr}}[A^\dagger d A\lambda_3]E_{\text{el}}$, with 
\begin{align}
\Delta c
&=c'\int {\text{d}}{\bf r}~\frac43\sin(F)\left(\frac{2F'}{r}+\frac{\sin(F)\cos(F)}{r^2}\right).
\end{align}
Since the relevant length scale of the soliton is $1/M_\chi$, a momentum expansion is not meaningful here. Hence \eqref{opd} is just one of the many contributions to $c$. The NDA guess $c\sim 4\pi c'/M_\chi\sim 4\pi f_\pi^2/M_\chi^2\sim N_c/(4\pi)$ is nevertheless expected to represent a reasonable estimate.\footnote{In a large $N_c$ expansion the tree exchange of $\eta'$ may result in contributions to the light mesons Lagrangian that at momenta $p^2<m_{\eta'}^2$ are of order $N_c^2$ rather than the usual ${\cal O}(N_c)$. Yet, such an enhancement is not a concern here because the relevant momentum flowing in the $ U$ field is set by $M_\chi\gg m_{\eta'}$.}

Given \eqref{EDMB}, our analysis is basically a carbon copy of the one presented in \cite{Adkins:1984cf}. The flavor-singlet component in $d$ does not contribute because of the unitarity of $A$. This is a consequence of the fact that ${\text{Tr}}[d]$ can only arise from a quark loop, which is subleading in the $1/N_c$ expansion. At leading order the baryon EDMs are a linear combination of operators of the form ${\text{Tr}}[A^\dagger \lambda_a A\lambda_k]$. The latter transform as octets of $SU(3)_V$ and triplets of $SU(2)_J$. If properly normalized, these can be identified as the irreducible octet representation of $SU(3)_V$:
\begin{align}
\frac12{\text{Tr}}[A^\dagger \lambda_a A\lambda_b]=D^{(8)}_{{\mathbb a},{\mathbb b}}(A).
\end{align}
Here ${\mathbb a}=(Y,I,I_3)$ is the index running through the components of the irreducible representation, i.e. the quantum numbers hypercharge, isospin and the third component of isospin. The same definition applies to ${\mathbb b}$. The operators we are interested in are ${\text{Tr}}[A^\dagger \lambda_8 A\lambda_3]$, for which we have ${\mathbb a}=(0,0,0)$, ${\mathbb b}=(0,1,0)$; then ${\text{Tr}}[A^\dagger \lambda_3 A\lambda_3]$, which corresponds to ${\mathbb a}=(0,1,0)$, ${\mathbb b}=(0,1,0)$; and finally ${\text{Tr}}[A^\dagger (\lambda_6-i\lambda_7) A\lambda_3]$, for which we have ${\mathbb a}=(-1,\frac12,\frac12)$, ${\mathbb b}=(0,1,0)$.

The third component of the dipole \eqref{EDMB} may now be written as
\begin{align}
{\bf d}_B
&= 2c\left\{
\frac{1}{2}(d_u-d_d)D^{(8)}_{(0,1,0),(0,1,0)}+\frac{1}{2\sqrt{3}}(d_u+d_d-2d_s)D^{(8)}_{(0,0,0),(0,1,0)}\right\}\\\no
&+2c\left\{\frac12d_{sd}D^{(8)}_{(-1,\frac12,\frac12),(0,1,0)}+\frac12d_{sd}^*D^{(8)}_{(1,-\frac12,\frac12),(0,1,0)}\right\}.
\end{align}
We will next focus on the neutron EDM. The dipoles of the other hadrons in the exact $SU(3)_V$ limit can be easily derived from that of the neutron using the results of Appendix \ref{app:chiralPT}. The neutron, with third component of the spin $J_3$, has a wavefunction given by \cite{Guadagnini:1983uv}
\ba\label{Psin}
\Psi_{n,J_3}(A)=\sqrt{8}D^{(8)}_{(1,\frac12,-\frac12),(1,\frac12,-J_3)},
\ea
and is normalized over the entire $SU(3)_V$ group manifold, i.e. $\int d\Omega\,\Psi_{n,J_3}^*\Psi_{n,J_3}=1$. The index ${\mathbb a}=(1,\frac12,-\frac12)$ is as expected, whereas ${\mathbb b}=(1,\frac12,-J_3)$ deserves an explanation. The first entry is related to the ambiguity $A\to Ae^{i\theta\lambda_8}$, the second is the eigenvalue of the spin squared operator, and the last entry is the third component of the baryon spin, which in \eqref{Psin} comes with an opposite sign compared to the associated physical state because rotations act by right multiplication on $A$.

Using the Clebsch-Gordan coefficients tabulated in \cite{deSwart:1963pdg} and \cite{McNamee:1964xq}, the neutron EDM becomes
\begin{align}\label{result}
\left.d_n\right|_{\text{soliton}}&=\int {\text{d}}\Omega\,\Psi_{n,+}^*{\bf d}_B\Psi_{n,+}\\\no
&=\int {\text{d}}\Omega\,\Psi_{n,+}^*\Psi_{n,+}\left\{c(d_u-d_d)D^{(8)}_{(0,1,0),(0,1,0)}+c\frac{1}{\sqrt{3}}(d_u+d_d-2d_s)D^{(8)}_{(0,0,0),(0,1,0)}\right\}\\\no
&=\frac{c}{5}\left[d_u-\frac43d_d+\frac13d_s\right].
\end{align}
In conclusion we have found 
\ba\label{ratios}
\frac{g_{Tn}^d}{g_{Tn}^u}=-\frac43,~~~~~~~\frac{g_{Tn}^s}{g_{Tn}^u}=\frac13,~~~~~~~(SU(3)_V{\text{-soliton}}).
\ea
The result is compatible with \eqref{largeNest}. The difference between the two is of order $10\%$ for $g_{Tn}^d/g_{Tn}^u$ and of order $1/N_c=30\%$ for $g_{Tn}^s/g_{Tn}^u$. The agreement stems from the fact that both descriptions rely on the same basic assumptions, namely large $N_c$ and $SU(3)_V$ flavor.

The soliton model prediction can also be extended to large $N_c$. The wavefunction \eqref{Psin} of the neutron is replaced by the one corresponding to the representation of $SU(3)_V$ shown in \eqref{normn}, and the matrix element \eqref{result} may be determined with the appropriate Clebsch-Gordan coefficients (see e.g. \cite{Cohen:2004ki}). That exercise has been carried out in \cite{Olness:1992zb} (for the proton). It is found that $g_{Tn}^d/g_{Tn}^u=-1+{\cal O}(1/N_c^2)$ and $g_{Tn}^s/g_{Tn}^u={\cal O}(1/N_c^2)$ in the limit $N_c\to\infty$, whereas of course one recovers \eqref{ratios} for $N_c=3$. The predictions found evaluating \eqref{result} are however affected at ${\cal O}(1/N_c)$ by operators with a single time derivative on $A$. As a consequence the soliton model predicts $g_{Tn}^s/g_{Tn}^u={\cal O}(1/N_c)$, consistently with what we found in Section \ref{sec:largeN}.

\subsection{Chromo-electric dipole moments at large $N_c$}
\label{sec:chromo}

One could repeat a large $N_c$ analysis similar to the one presented above for quark EDMs also for hypothetical chromo-electric dipole moments of the quarks. The latter are defined by the CP-violating operator
\begin{align}\label{Lag1}
\delta{\cal L}_{\Lambda_{\text{UV}}}
&=-\left(\frac{\tilde d_{ij}}{2}\overline{q}_ii\sigma^{\mu\nu}\gamma_5T^Aq_j\,G^A_{\mu\nu}+{\text{hc}}\right),
\end{align}
with
\begin{align}\label{Cgen}
\tilde d&=\left(
\begin{matrix} \tilde d_u & &\\ & \tilde d_d & \tilde d_{sd}^*\\ & \tilde d_{sd} & \tilde d_s\end{matrix}\right).
\end{align}
At first non-trivial order in the new couplings, similarly to \eqref{nEDM} one can write 
\ba\label{CEDM}
d_n=\tilde d_{u}\tilde g_{Tn}^u+\tilde d_{d}\tilde g_{Tn}^d+\tilde d_{s}\tilde g_{Tn}^s. 
\ea
There is an important difference compared to quark EDMs, though. The spurion $\tilde d$ impacts the meson potential more significantly than $d$, as the latter appears in the vacuum energy multiplied by a suppression due to the electric charge whereas the former does not. This has two important implications. First, $\tilde d$ introduces novel meson-baryon CP-violating couplings that should be taken into account in the chiral expansion. Second, in the presence of a dynamical relaxation of $\bar\theta$, the coupling $\tilde d$ would induce a residual $\bar\theta_{\text{eff}}$ that results in a contribution to $d_n$ that is of the same parametric order as the one, always present, coming directly from $\tilde d$. As a result, the final expression of $\tilde g_{Tn}^i$ depends on whether an axion exists or not. This is not the case for the quark EDMs discussed here, as already mentioned below \eqref{dgen}.

Nevertheless, the leading order direct contributions predicted by the large $N_c$ counting (and similarly the soliton model of Appendix \ref{app:soliton}) are essentially analogous to those derived for the quark EDMs provided one replaces 
\ba
d\rightsquigarrow \frac{e}{4\pi}Q\tilde d, 
\ea
with $\rightsquigarrow$ indicating that the mapping includes an unknown number of order unity and $Q={\text{diag}}(\frac23,-\frac13,-\frac13)$ is the quark charge matrix. The result is that, at leading order, $\tilde g_{Tn}^d/\tilde g_{Tn}^u=-\frac12g_{Tn}^d/g_{Tn}^u$ and $\tilde g_{Tn}^s/\tilde g_{Tn}^u=-\frac12g_{Tn}^s/g_{Tn}^u$.

\section{Baryon EDMs up to ${\cal O}(m_sd)$}
\label{app:chiralPT}

Next to leading order calculations in chiral perturbation theory are standard since many decades by now. Nevertheless, for completeness and transparency we present details of the salient steps of our analysis. An analogous 1-loop calculation for the two-flavor case has been performed for example in \cite{deVries:2010ah}.

It is convenient to introduce the notation $\Psi_{ij}=[\tilde\lambda_{a'}]_{ij}\Psi_{a'}$, adopting the following conventional ordering
\begin{align}
\Psi_{a'}=\Sigma_+,~\Sigma_-,~\Sigma_0,~p,~n,~\Xi_-,~\Xi_0,~\Lambda, 
\end{align}
where $\tilde\lambda_{a'}$ can be readily seen from \eqref{PsiB}. With this notation the baryon couplings become traces of $\tilde\lambda_{a'}$'s and $T_a$'s, the latter being the normalized flavor matrices $T_a=\lambda_a/2$. As explained in the text, at the order we are working all the baryon octet components have the same mass.

Using the notation ${\cal L}_{\text{EFT}}\supset -\frac{i}{2}d_{a'b'}\overline{\Psi}_{a'}\sigma^{\mu\nu}\gamma_5\Psi_{b'}F_{\mu\nu}$, where $d_{a'b'}=d_{b'a'}^*$, and adopting Dimensional Regularization, the formal expression of the baryon octet dipoles, up to NLO, reads 
\begin{align}\label{NLOdimreg}
d^{\text{NLO}}_{a'b'}
&=d^{\text{LO}}_{a'b'}\\\no
&+A^K_{a'b'}\frac{M_K^2}{16\pi^2f_\pi^2}\ln\frac{\mu^2}{M_K^2}+A^\eta_{a'b'}\frac{M_\eta^2}{16\pi^2f_\pi^2}\ln\frac{\mu^2}{M_\eta^2}\\\no
&+B^K_{a'b'}\frac{M_K^2}{16\pi^2f_\pi^2}.
\end{align}
The leading order expression can be directly seen from \eqref{dip} and is given by
\ba\label{BLO}
d^{\text{LO}}_{a'b'}&={\text{Tr}}\left[\kappa_1 d\tilde\lambda_{a'}^\dagger \tilde\lambda_{b'}+\kappa_2\tilde\lambda_{a'}^\dagger d\tilde\lambda_{b'}+\kappa_3{\text{Tr}}[d]\tilde\lambda_{a'}^\dagger \tilde\lambda_{b'}\right].
\ea
Explicitly, the LO dipoles are
\begin{align}\label{SU3EDM1}
d^{\text{LO}}_{nn}&=\kappa_1 d_s+\kappa_2 d_d+\kappa_3(d_u+d_d+d_s)\\\no
d^{\text{LO}}_{pp}&=\kappa_1 d_s+\kappa_2 d_u+\kappa_3(d_u+d_d+d_s)\\\no
d^{\text{LO}}_{\Xi_-\Xi_-}&=\kappa_1 d_u+\kappa_2 d_s+\kappa_3(d_u+d_d+d_s)\\\no
d^{\text{LO}}_{\Xi_0\Xi_0}&=\kappa_1 d_d+\kappa_2 d_s+\kappa_3(d_u+d_d+d_s)\\\no
d^{\text{LO}}_{\Sigma_-\Sigma_-}&=\kappa_1 d_u+\kappa_2 d_d+\kappa_3(d_u+d_d+d_s)\\\no
d^{\text{LO}}_{\Sigma_+\Sigma_+}&=\kappa_1 d_d+\kappa_2 d_u+\kappa_3(d_u+d_d+d_s)\\\no
d^{\text{LO}}_{\Sigma_0\Sigma_0}&=\kappa_1 \left(\frac{1}{2}d_u+\frac{1}{2}d_d\right)+\kappa_2 \left(\frac{1}{2}d_u+\frac{1}{2}d_d\right)+\kappa_3(d_u+d_d+d_s)\\\no
d^{\text{LO}}_{\Lambda\Lambda}&=\kappa_1 \left(\frac{1}{6}d_u+\frac{1}{6}d_d+\frac{2}{3}d_s\right)+\kappa_2 \left(\frac{1}{6}d_u+\frac{1}{6}d_d+\frac{2}{3}d_s\right)+\kappa_3(d_u+d_d+d_s)
\end{align}
and the transition dipoles are
\begin{align}\label{SU3EDM2}
d^{\text{LO}}_{\Lambda\Sigma_0}&=\kappa_1 \left(\frac{\sqrt{3}}{{6}}(d_u-d_d)\right)+\kappa_2 \left(\frac{\sqrt{3}}{{6}}(d_u-d_d)\right)\\\no
d^{\text{LO}}_{{\Sigma_+}p}&=\kappa_1 d_{sd}\\\no
d^{\text{LO}}_{\Sigma_0n}&=\kappa_1 \left(-\frac{1}{\sqrt2}d_{sd}\right)\\\no
d^{\text{LO}}_{\Lambda n}&=\kappa_1 \left(\frac{1}{\sqrt{6}}d_{sd}\right)+\kappa_2 \left(-\frac{2}{\sqrt6}d_{sd}\right)\\\no
d^{\text{LO}}_{\Xi_0\Lambda}&=\kappa_1 \left(-\frac{2}{\sqrt{6}}d_{sd}\right)+\kappa_2 \left(\frac{1}{\sqrt6}d_{sd}\right)\\\no
d^{\text{LO}}_{{\Xi_-}\Sigma_-}&=\kappa_2 d_{sd}\\\no
d^{\text{LO}}_{\Xi_0\Sigma_0}&=\kappa_2 \left(-\frac{1}{\sqrt2}d_{sd}\right).
\end{align}
In the paper for simplicity we wrote $d_{nn}=d_n$. The relation to the notation introduced in \eqref{nEDM} is given in \eqref{LOcoeff}.

The matrixes $A^{K,\eta}_{a'b'}$ in \eqref{NLOdimreg} are obtained computing the diagrams in Fig. \ref{Fig}, and are fully determined by the leading order Lagrangian:
\begin{align}\label{NLOresult}
A^K_{a'b'}&=\sum_{a=4}^7\left\{3{\cal T}^a_{a'm'}{\cal T}^a_{n'b'}d^{\text{LO}}_{a'b'}+{\cal D}_{a'b'}^{aa}-{\cal T}^a_{a'm'}d^{\text{LO}}_{m'n'}{\cal T}^a_{n'b'}\right\}\\\no
A^\eta_{a'b'}&=3{\cal T}^8_{a'm'}{\cal T}^8_{n'b'}d^{\text{LO}}_{a'b'}+{\cal D}_{a'b'}^{88}-{\cal T}^8_{a'm'}d^{\text{LO}}_{m'n'}{\cal T}^8_{n'b'}.
\end{align}
The structure $3{\cal T}^a_{a'm'}{\cal T}^a_{n'b'}d^{\text{LO}}_{a'b'}$ arises from wavefunction renormalization, ${\cal T}^a_{a'm'}d^{\text{LO}}_{m'n'}{\cal T}^a_{n'b'}$ comes from the first diagram in Fig. \ref{Fig}, whereas ${\cal D}_{a'b'}^{aa}$ from the second. The third diagram identically vanishes in Dimensional Regularization because of the non-relativistic condition ${\slashed{v}}\Psi=0$. Hence, the CP-violating couplings in the second line of ${\cal L}_{\text{dip}}$ do not contribute (at 1-loop) in this scheme. Eqs. \eqref{NLOresult} depend on the $\Psi^2\Pi$ couplings in \eqref{LO}, which read
\ba\label{BT}
{\cal T}^a_{a'b'}&={\text{Tr}}\left[(D+F)\tilde\lambda_{a'}^\dagger T^a\tilde\lambda_{b'}+(D-F)\tilde\lambda_{a'}^\dagger \tilde\lambda_{b'}T^a\right],
\ea
as well as the $\Psi^2\Pi^2$ couplings in the first line of \eqref{dip}:
\begin{align}\label{def2}
{\cal D}_{a'b'}^{ab}&=\kappa_1{\text{Tr}}\left\{[T^adT^b+\frac12 T^aT^bd+\frac12 dT^aT^b]\tilde\lambda_{a'}^\dagger \tilde\lambda_{b'}\right\}\\\no
&+\kappa_2{\text{Tr}}\left\{\tilde\lambda_{a'}^\dagger [T^adT^b+\frac12 T^aT^bd+\frac12 dT^aT^b]\tilde\lambda_{b'}\right\}\\\no
&+\kappa_3{\text{Tr}}[T^adT^b+\frac12 T^aT^bd+\frac12 dT^aT^b]{\text{Tr}}\left\{\tilde\lambda_{a'}^\dagger \tilde\lambda_{b'}\right\}.
\end{align}
Finally, $B^{K}_{a'b'}$ in \eqref{NLOdimreg} include the effect of unknown counterterms of ${\cal O}({m}d)$ as well as of the loops involving the baryon decuplet (their dipoles and mixed octet-decuplet dipoles) and heavier states. The complete list of counterterms is quite long, but fortunately we do not need to display it here. Since our main focus is the neutron EDM, it suffices to say that each of the parameters $g_{Tn}^{u,d,s}$ of \eqref{nEDM} is associated to an independent counterterm of order $M_K^2/\Lambda_\chi^2$. The contribution of the decuplet loops is expected to be at most of the same order since the decuplet-octet mass splitting is smaller than $M_K$. We thus write the counterterm of the neutron as $B^{K}_{nn}=B^{K}_{nn,u}d_u+B^{K}_{nn,d}d_d+B^{K}_{nn,s}d_s$ with $B^{K}_{nn,i}$ of order unity.

Eqs. \eqref{NLOdimreg} and \eqref{NLOresult} summarize the main result of this appendix. Explicit NLO expressions for the baryons EDMs can be derived from the definitions given in \eqref{BLO}, \eqref{NLOresult}, \eqref{BT}, and \eqref{def2}. We quote only the expression for the neutron. Using the convenient notation \eqref{LOcoeff}, and in terms of couplings renormalized at the scale $\mu=M_n$, we have
\begin{align}\label{here1}
g_{Tn,\text{NLO}}^u
&=g_{Tn,\text{LO}}^u\\\no
&+\frac{M_K^2}{16\pi^2f_\pi^2}\ln\frac{M_n^2}{M_K^2}\left(\begin{matrix} 
\frac12+\frac{47}{18}D^2-3DF+\frac72F^2\\-\frac29D^2\\ \frac12-\frac{13}{18}D^2+DF-\frac12 F^2
\end{matrix}\right)^t\left(\begin{matrix} g_{Tn,\text{LO}}^u\\ g_{Tn,\text{LO}}^d\\ g_{Tn,\text{LO}}^s\end{matrix}\right)\\\no
&+\frac{M_\eta^2}{16\pi^2f_\pi^2}\ln\frac{M_n^2}{M_\eta^2}\left(\frac16+\frac{1}{6}(D-3F)^2\right)g_{Tn,\text{LO}}^u+B_{nn,u}^K\frac{M_K^2}{16\pi^2f_\pi^2}\\\label{here2}
g_{Tn,\text{NLO}}^d
&=g_{Tn,\text{LO}}^d\\\no
&+\frac{M_K^2}{16\pi^2f_\pi^2}\ln\frac{M_n^2}{M_K^2}\left(\begin{matrix} 
-\frac29 D(D+3F) \\ \frac12+\frac{35}{18}D^2-\frac{5}{3} DF+\frac72F^2 \\ \frac12-\frac{1}{18}(D-3F)^2
\end{matrix}\right)^t\left(\begin{matrix} g_{Tn,\text{LO}}^u\\ g_{Tn,\text{LO}}^d\\ g_{Tn,\text{LO}}^s\end{matrix}\right)\\\no
&+\frac{M_\eta^2}{16\pi^2f_\pi^2}\ln\frac{M_n^2}{M_\eta^2}\left(\frac16+\frac{1}{6}(D-3F)^2\right)g_{Tn,\text{LO}}^d+B_{nn,d}^K\frac{M_K^2}{16\pi^2f_\pi^2}\\\label{here3}
g_{Tn,\text{NLO}}^s
&=g_{Tn,\text{LO}}^s\\\no
&+\frac{M_K^2}{16\pi^2f_\pi^2}\ln\frac{M_n^2}{M_K^2}\left(\begin{matrix} 
\frac12-\frac{13}{18}D^2+\frac{5}{3}DF-\frac12 F^2 \\ \frac12-\frac{1}{18}(D+3F)^2 \\ 1+\frac{22}{9}D^2-\frac{10}{3}DF+4F^2
\end{matrix}\right)^t\left(\begin{matrix} g_{Tn,\text{LO}}^u\\ g_{Tn,\text{LO}}^d\\ g_{Tn,\text{LO}}^s\end{matrix}\right)\\\no
&+\frac{M_\eta^2}{16\pi^2f_\pi^2}\ln\frac{M_n^2}{M_\eta^2}\left(\frac23+\frac{1}{6}(D-3F)^2\right)g_{Tn,\text{LO}}^s+B_{nn,s}^K\frac{M_K^2}{16\pi^2f_\pi^2}.
\end{align}
The analogous expressions for the proton are obtained identifying $g_T^u=g_{Tn}^d$, $g_T^d=g_{Tn}^u$, $g_T^s=g_{Tn}^s$.

\end{document}